\documentclass[epj-spec]{svjour}

\usepackage{graphicx}
\usepackage{bm}
\usepackage{amssymb}

\newcommand{\bd}{\begin{document}}
\newcommand{\ed}{\end{document}}
\newcommand{\nid}{\noindent}
\newcommand{\beq}{\begin{equation}} 
\newcommand{\eeq}{\end{equation}}
\newcommand{\bea}{\begin{eqnarray}} 
\newcommand{\eea}{\end{eqnarray}}
\newcommand{\Di}{\displaystyle}


\def \du{\dot{u}}
\def \dV{\dot{V}} 
\def \EE{\mbox{E}}
\def \Var{\mbox{Var}}
\def \Cov{\mbox{Cov}}
\newcommand{\DD}{{\rm I\hspace{-0.55ex}D}}
\newcommand{\NN}{\mathbb{N}}
\newcommand{\PP}{{\rm I\hspace{-0.55ex}P}}
\newcommand{\QQ}{\mathbb{Q}}
\newcommand\bbbr{{\sf I\!R}}
\newcommand{\RR}{\mathbb{\bbbr}}
\newcommand{\UN}{{\rm 1\hspace{-0.55ex}I}}
\newcommand{\ZZ}{\mathbb{Z}}

\newcommand{\ds}{\displaystyle}
\newcommand{\ba}{\begin{array}{l}}
\newcommand{\ea}{\end{array}}
\newcommand{\sys}[2]{\be \left\{ \ba \ds #1 \ea \right. \label{#2} \ee}

\newcommand{\dx}{\frac{dx}{dt}}
\newcommand{\dep}{\partial}
\newcommand{\ddt}[1]{\frac{d #1}{dt}}
\newcommand{\depdt}[1]{\frac{\partial #1}{\partial t}}
\newcommand{\II}{\parallel}
\newcommand{\norm}[1]{\parallel #1 \parallel}
\newcommand{\abs}[1]{\mid #1 \mid}
\newcommand{\ps}[2]{ \left( #1 \mid #2 \right)}

\newcommand{\ra}{\rightarrow}
\newcommand{\limt}{\lim_{t \rightarrow \infty} }
\newcommand{\Dfxo}{\nabla_{x_0}f}
\newcommand{\DEw}{\nabla_{w}\mc{E}}
\newcommand{\Spfxo}{\mathrm{Sp}\left(\nabla_{x_0}f \right)}
\newcommand{\Tr}{\mathrm{Tr}}

\def \imp{\Rightarrow}
\def \va{variable al\'eatoire }
\def \veca{vecteur al\'eatoire }
\def \*{\textrm{o}}
\def \.{\bullet}
\def \x{\tilde{x}}
\def \*{^\star}

\def \ps{\textrm{ p.s. }}
\newcommand{\tdg}{\theta\rightarrow\partial\Gamma}
\newcommand{\tdi}{\parallel\theta\parallel\rightarrow\infty}

\newcommand{\deq}{\stackrel {\rm def}{=}}
\newtheorem{theo}{Theorem}
\newtheorem{prop}[theo]{Proposition}
\newtheorem{lemme}[theo]{Lemma}
\newtheorem{defi}{Definition}
\newtheorem{rem}{Remark}
\newtheorem{coro}[theo]{Corollary}
\newtheorem{exemple}{Example}
\newtheorem{hypo}[theo]{Hypothesis}

\newcommand{\ex}{\textbf{Exemple:}\\}
\newcommand{\Ex}[1]{\begin{exemple} {\bf #1} \end{exemple}}
\newcommand{\Rem}[1]{\begin{rem} {\bf #1} \end{rem}}
\newcommand{\pr}{P{\small ROOF} : }
\newcommand{\prlemme}{P{\small ROOF OF LEMMA}: }
\newcommand{\prtheo}{P{\small ROOF OF THEOREM}: }
\newcommand{\cqfd}{\hfill\rule{0.25cm}{0.25cm}}


\def \"{\textrm}

\def \al{\alpha}
\def \bt{\beta}
\def \del{\delta}
\def \ga{\gamma}
\def \eps{\epsilon}
\def \la{\lambda}
\def \ta{\theta }
\def \om{\omega }
\def \sig{\sigma}
\def \oups{\bar{J}}
\def \ups{J^2}
\def \e {\varepsilon}
\def \vfi{\varphi}
\def \vta{\vartheta}

\def \Del{\Delta}
\def \Ga{\Gamma}
\def \La{\Lambda}
\def \Ta{\Theta}
\def \Om{\Omega}
\def \Sig{\Sigma}

\def \mc{\mathcal}
\def \ul{\underline}
\def \ol{\overline}

\def \a{\mc{a}}
\def \b{\mc{b}}
\def \d{\mc{d}}
\def \e{\mc{e}}
\def \f{\mc{f}}
\def \g{\mc{g}}
\def \h{\mc{h}}
\def \k{\mc{k}}
\def \l{\mc{l}}
\def \m{\mc{m}}
\def \n{\mc{n}}
\def \o{\mc{o}}
\def \p{\mc{p}}
\def \r{\mc{r}}
\def \s{\mc{s}}
\def \t{\mc{t}}
\def \u{\mc{u}}
\def \v{\mc{v}}

\def \A{\mc{A}}
\def \B{\mc{B}}
\def \C{\mc{C}}
\def \D{\mc{D}}
\def \E{\mc{E}}
\def \F{\mc{F}}
\def \G{\mc{G}}
\def \H{\mc{H}}
\def \J{\mc{J}}
\def \K{\mc{K}}
\def \L{\mc{L}}
\def \M{\mc{M}}
\def \N{\mc{N}}
\def \O{\mc{O}}
\def \Pp{\mc{P}}
\def \R{\mc{R}}
\def \Ss{\mc{S}}
\def \T{\mc{T}}
\def \U{\mc{U}}
\def \V{\mc{V}}
\def \W{\mc{W}}
\def \X{\mc{X}}
\def \Y{\mc{Y}}

\def \Gav{\mathbf{\Gamma}}
\def \muv{\mathbf{\mu}}
\def \piv{\mathbf{\pi}}
\def \Kv{\mathbf{K}}
\def \Av{\mathbf{A}}
\def \Cv{\mathbf{C}}
\def \uv{\mathbf{u}}
\def \xv{\mathbf{x}}
\def \Xv{\mathbf{X}}
\def \Yv{\mathbf{Y}}


\newcommand{\Fig}[3]{
\begin{figure}[!htb]
\centering
\epsfig{file =3D \adresseFIGURES #1.eps, width=3D #2 \linewidth}
\caption{#3}
\mlabf{#1}
\end{figure}
}

\bd

\title{Random Recurrent Neural Networks Dynamics.}
\author{
 M. Samuelides \inst{1} \and B. Cessac \inst{2,3,4}
}

\institute{ Ecole Nationale Sup\'erieure de l'A\'eronautique et de l'espace and ONERA/DTIM,
2 Av. E. Belin, 31055 Toulouse, France.
\and INRIA, 2004 Route des Lucioles, 06902 Sophia-Antipolis, France.
\and INLN, 1361, Route des Lucioles, 06560 Valbonne, France.
\and Universit\'e de Nice, Parc Valrose, 06000 Nice, France.
}

\abstract{
This paper is a review dealing with  the study of large size random recurrent
neural networks. The connection weights are varying according to a
probability law and it is possible to predict the network dynamics
at a macroscopic scale using an averaging principle. After a first
introductory section, the section \ref{DynRRNN} reviews the various models from
the points of view of the single neuron dynamics and of the global
network dynamics. A summary of notations is presented, which is quite
helpful for the sequel. In section \ref{DMFT}, mean-field dynamics is developed.
 The probability distribution characterizing global dynamics is computed.
In section \ref{MFAnalog}, some applications of mean-field theory to the prediction of
chaotic regime for Analog Formal Random Recurrent Neural Networks
(AFRRNN) are displayed. The case of AFRRNN with an homogeneous population of neurons
 is studied in section \ref{MFDHN}. Then, a two-population model is studied in section \ref{MFD2Pop}.
The occurrence of a cyclo-stationary chaos is displayed using the
results of \cite{Dauce01}. In section \ref{MFOscIF}, an insight of the application of
mean-field theory to IF networks is given using the results of
\cite{BrunelHakim99}.
}


\maketitle

\section{Introduction}\label{Intro}


Recurrent neural networks were introduced to improve
biological plausibility of artificial neural networks such as
perceptrons since they display internal dynamics. They are useful to
implement associative recall. The first models were endowed with
symmetric connexion weights which induced relaxation dynamics and
equilibrium states ~\cite{Hopfield82}. Asymmetric
connexion weights were introduced later on, enabling the observation of
complex dynamics and chaotic attractors. The role of chaos in
cognitive functions was first discussed by W.Freeman and C.Skarda
in seminal papers such as ~\cite{SkardaFreeman90}. The practical
importance of such dynamics is due to the use of on-line Hebbian
learning to store dynamical patterns. For a review see for
instance~\cite{GuillotDauce02}.

The nature of the dynamics depends on the connexion weights. When
considering large sized neural networks, it is impossible to study
the dynamics as a function of the 
 many parameters involved in the network dynamics: parameters defining
the state of the neuron such as the Sodium, Potassium conductibility
in Hodgkin-Huxley models; parameters defining the structure of the
synapses; parameters attached to the environment; external inputs;
etc \dots. 
 One may
consider that the connexion weights share few values but this does not
allow to study the effect of the variability. Henceforth, one often considers
\textit{random} models where the connexion weights form a random sample of a
probability distribution. These models are called \textit{"Random Recurrent
Neural  Networks"} (RRNN). 

In this context, the parameters of interest
are those defining the probability distribution,  i.e. the statistical parameters
(introduced as ``macroscopic parameters'' in the first paper of this review, refereed, from now on,
as paper I).
Then 
a study of the dynamics in terms of relevant dynamical quantities,
called \textit{order parameters\footnote{This terminology comes
from statistical physics and was introduced by the physicists L. Landau. 
Prominent examples
of order parameters are the magnetization in the Ising model or
the Edwards-Anderson parameter in spin-glasses.} } 
can be performed via 
\textit{"Dynamic Mean-Field Equations"} (DMFE).
 This terminology and method is inherited from statistical
physics and quantum field theory, though is has to be adapted to the present context.
Mean-Field Equations were introduced for neural networks by
Amari~\cite{Amari77} and later on by Crisanti and Sompolinsky~\cite{Sompo88}. 
Their 
results were extended in ~\cite{Cessac94} and  proved  in a
rigorous way in~\cite{Moynot02}. In ~\cite{Moynot02} the Authors
used a \textbf{"Large deviation Principle"} (LDP) coming from 
rigorous statistical mechanics~\cite{Benarous95}.
Simultaneously,
mean-field equations were successfully used to predict the dynamics
of spiking recurrent neural networks \cite{BrunelHakim99,BMS}.\\

This paper intends to
provide a bridge between the detailed computation of the asymptotic
regime and the rigorous aspects of MFE theory.
We shall introduce the mathematical basis of dynamic mean field theory
in sections \ref{DynRRNN},\ref{DMFT}, and apply it to several prominent examples.
 Note however that our approach is different from the standard ones based either on
the computation of a functional generating the cumulants, or using an
ad hoc approximation replacing the sum of incoming synaptic potentials
by a Gaussian random variable. Instead we use large deviations techniques
(detailed in the appendix). They have the advantage to be rigorous and
they allows to prove convergence results stronger than the usual techniques.

In section \ref{DynRRNN}, the various models are stated from the points of view
of the single neuron dynamics and of the global network dynamics. A
summary of notations is presented, which is quite helpful for the
sequel. In section \ref{DMFT} mean-field dynamics is developed.
 The probability distribution characterizing global dynamics is computed.
 The mathematical tools which are
used there are detailed (without any proof) in appendix.
In section \ref{MFAnalog}, some applications of mean-field theory to the
prediction of chaotic regime for analog formal random recurrent
neural networks (AFRRNN) are displayed. The dynamical equation of
homogeneous AFRRNN,  which is studied in paper I, is derived from
the random network model in section \ref{MFDHN}. Moreover a two-population
model is studied in section \ref{MFD2Pop} and the occurrence of a
cyclo-stationary chaos is displayed using the results of
\cite{Dauce01}. In section \ref{MFOscIF}, an insight of the application of
mean-field theory to IF networks is given using the results of
\cite{BrunelHakim99}. The model of this section is a
continuous-time model following the authors of the original
paper. Hence the theoretical framework of the beginning of the
paper has to be enlarged to support this extension of mean-field
theory and this work has still to be done. However, we sketch a
parallel between the two models to induce further research.

\section{Dynamics of Random Recurrent Neural Networks.}\label{DynRRNN}

\subsection{Defining dynamic state variables} \label{SDef}

The state of an individual neuron $i$ at time $t$ is described by
an instantaneous individual variable, the \textit{membrane
potential} $u_i(t)$. In  stochastic models, such as the ones
considered here,   all variables (including the $u_i(t)$'s) are (real)
 random variables\footnote{Following the
standard physicist's habit, the random variables won't be noted by
capital letters.} which takes
their values in $\RR$. In this section, we consider discrete time dynamics 
and restrict ourselves to
finite time-horizon, i.e. we consider time $t$ as an integer
belonging to time interval$\{0,1,...,T\}$ where $T$
is finite.  Thus  an individual state trajectory
$u_i \deq (u_i(t))_{t\in\{0,1,...,T\}}$ takes its value in
$\F = \RR^{\{0,1,...,T\}}$. Though one is also
 interested in long-time behaviour and stationary regime (if
any), rigorous proofs of convergence of large-size networks only
exist for finite time. 

We shall  study the probability distribution of 
\textit{trajectories} instead of the probability of an instantaneous state. 
Actually, the later can be easily obtained from the former.
The
second order moments of an individual trajectory $u_i$ are its
expectation $\EE(u_i)\in\F$ and its covariance matrix
$\Cov(u_i)\in\F\otimes\F$.\\

Our aim is to study the coupled dynamics of $N$ interacting neurons
that constitute a neural  network. $N$ is the \textit{size} of the
neural network. The global state trajectory of
$u \deq (u_i)_{i\in\{1...,N\}}$ is a random vector in $\F^N$. The
probability distribution\footnote{This term is defined in Appendix,
definition \ref{law.def}} of the random vector $u$, denoted
by $Q_N$  depends on $N$. We shall compute it for various neuron models in this
section. We shall first focus to the case of
homogeneous neural networks then more realistic cases such as several
population networks will be considered.  \\

As it was detailed in paper I, the dynamics of the
neuron models that we study here depends on three crucial points.
\begin{itemize}
\item  how does the neuron activation depend on the membrane
potential,
\item how do the other neurons contribute to the synaptic
potential which summarizes completely the influence of the network
onto the target neuron,
\item how is the synaptic potential used to update the
membrane potential.
\end{itemize}

We shall now detail these points in the models that are
considered further on.

\subsection{Spike firing modeling}\label{SSPikeFire}

As introduced in paper I, it is considered that the neuron is active and emits a spike whenever
its membrane potential exceeds the \textit{activation threshold}. So the
neuron $i$ is active at time $t$ when $u_i(t)\geq\ta$ where $\ta$
is the neuron activation threshold. We consider here that
$\ta$ is  constant during the evolution and  is the same for all neurons.
Actually this hypothesis may be relaxed and random thresholds may be
considered but the notation and the framework of dynamical study
would be more complicated (see \cite{Moynot02}).\\

For spiking neuron models we define an \textit{activation variable}
$x_i(t)$ which is equal to 1 if neuron $i$ emits a spike at time
$t$ and to 0 otherwise. Hence we have
\begin{equation}\label{transfer0.eq}
x_i(t)=f[u_i(t)-\ta]
\end{equation}
where $f$, called the
\textit{transfer function} of the neuron  is here the Heaviside function. Actually, to alleviate
notations, we shift $u_i$ of $\ta$ and that allows to replace
equation (\ref{transfer0.eq}) by equation
\begin{equation}\label{transfer.eq}
x_i(t)=f[u_i(t)]
\end{equation}
The threshold will be further on taken into account in the updating
equation.\\

Two spiking neuron models are considered here, the
\textit{Binary Formal  neuron (BF)} which is the original model of
Mac Culloch and Pitts \cite{Mcculloch43} and the \textit{Integrate
and Fire neuron (IF)} which is generally used nowadays to model
dynamics of large spiking neural networks \cite{Gerstner02}.\\

In these models, the neuron activation takes generally two values:
$0$ and $1$.  This is true for most models of neurons.
However, it was preferred in a lot of research works to take into
account the average firing rate of the model instead of the detailed
instant of firing (see section 5.1, paper I). This point of view simplifies the model as it
deals with smooth functions easier to handle from a
mathematical point of view. In this case,
equation (\ref{transfer.eq}) is still valid but $x_i(t)$ takes its
values in the interval $[0,1]$ and the transfer function $f$ is a
smooth sigmoid function for instance
$f(x)=\displaystyle{\frac{e^x}{1+e^x}}$. Since the activation of
the neuron is represented by a real value that varies continuously,
the model is called \textit{Analog Formal neuron (AF)}.

AF model is still widely dominant when Artificial Neural Networks
are considered for applications since gradient are easy to compute.
For biological purpose, it was widely believed that the relevant
information was stored in the firing rate; in that case more
precise modeling would not be so useful, at least from a functional
point of view. 

The three models are studied in that chapter and we
attempt to give a unified presentation of mean-field equation for
these three models.
Note that, in the sequel, the \textit{state} of the neuron will be  the membrane
potential $u_i(t)$ and not the activation $x_i(t)$. This is due to
the updating definition in the IF model. We shall return to that point
later on.

\subsection{The synaptic potential of RRNN}\label{SSynpot}

The spikes are used to transmit information to other neurons through
the synapses. We shall adopt here the very rough, but classical
description of the synapse. Namely, the synaptic connexion from neuron
$j$ to neuron $i$ is denoted by $J_{ij}$. It can be positive
(excitatory) or negative (inhibitory).
Let us denote by $\J=(J_{ij})$ the matrix of
synaptic weights.  At time $0$, the dynamical system is
initialized and the synaptic potentials are set to zero\footnote{This may
for example corresponds to a rest state set to zero without loss
of generality (since its corresponds to changing the voltage reference
 for the membrane
potential, see paper I).}.
 
The
\textit{synaptic  potential} of neuron $i$ of a network of $N$
neurons at time $t+1$ is expressed\footnote{See section 5 in paper I.} as a
 function of $\J$ and
$u(t)\in\RR^N$ by
\begin{equation}\label{synapticdyn.eq}
v_i(\J,u)(t+1)=\sum_{j=1}^N J_{ij}x_j(t)
=\sum_{j=1}^N J_{ij}f[u_j(t)]
\end{equation}

\noindent (The notation $v_i(\J,u)(t+1)$ will be explained below).\\

As discussed  in the introduction, we consider here
random models where the connexion weights form a random sample of a
probability distribution (\textit{"Random Recurrent
Neural  Networks"}(RRNN)). In that case, the parameters of interest
are the statistical parameters defining the probability distribution.
A standard example considered in this paper is  Gaussian connexion weights.
In this case, the statistical  parameters are denoted by\footnote{We use here the same notation
as in paper I. Recall that the scaling with $\frac{1}{N}$ allows
to have a synaptic potential whose mean and variance are independent of $N$.}
$\oups$ and $\ups$ so that  $\J$ is
a normal random matrix with
\textit{independent} components distributed according to the normal law  $\N(\frac{\oups}{N},\frac{\ups}{N})$. 

Note that the assumption of independence is crucial in the
approach described below. Unfortunately, the more realistic
case where correlations between the $J_{ij}$'s exist
(e.g. after Hebbian learning) is, currently,
out of reach for all the mean-field methods that we know.
We shall first consider Gaussian synaptic synaptic, but we shall extend later
on  the RRNN model properties to a more
general setting where the weights are non Gaussian and depend on
the neuron class in a several population model like
in~\cite{Dauce01}.\\

We have already dealt with the dynamical properties of
RRNN such as (\ref{synapticdyn.eq}) in paper I,
considered from the dynamical system point of view,
where we fix a realization of the $J_{ij}$'s and
consider the evolution of trajectories of this dynamical 
system. Then, we have averaged over the $J_{ij}$ distribution
in order to get informations about the evolution of averaged
quantities.  
In the present paper  we shall start with a complementary
point of view. Namely, assume that we \textit{fix the trajectory $u_i$ of each neuron}
(resp. we fix the trajectory $u$ of the network).
Then, at each time step the variable $\sum_{j=1}^N J_{ij}f[u_j(t)]$
is a Gaussian random variable whose probability distribution is induced
by the distribution of the $J_{ij}$'s. Of course, this distribution
depends on the trajectory (for example 
$E[\sum_{j=1}^N J_{ij}f[u_j(t)]]=\frac{\oups}{N}\sum_{j=1}^N f[u_j(t)]$).
To emphasize this dependence  we shall
denote by  $v_i(\J,u)=(v_i(\J,u)(t))\in\F$ the trajectory of the synaptic potential
as in (\ref{synapticdyn.eq}).

With this line of reasoning one can show that the $v_i(.,u)$ are
(conditionally on $u$)
Gaussian identically distributed and \textit{independent} random vectors in
$\F$ (see appendix, proposition (\ref{GaussLin.prop}). The distribution
of  $v_i$ is therefore\footnote{This distribution actually does not depend on $i$ since
the $v_i$'s are identically distributed.} defined by its mean $m_u$ and its covariance
matrix $c_u$ (depending on $u$).

We have
\begin{equation}
m_u(t+1)=\frac{\oups}{N}\sum_{j=1}^N f[u_j(t)]
\end{equation}
and
\begin{equation}
c_u(s+1,t+1)=\frac{\ups}{N}\sum_{j=1}^N f[u_j(s)]f[u_j(t)]
\end{equation}

Notice that these quantities, called \textit{order parameters} in the sequel, are invariant by any permutation of
the neuron membrane potentials. 
Actually, they depend only on
the \textit{empirical distribution}\footnote{This concept is
introduced within more details in the appendix, definition (\ref{empiric.def})}
$\mu_u$, associated to $u$. 

\begin{defi} The empirical measure is an application from $\F^N$ to
$\Pp(F)$, the set of probability measures on $\F$. It is defined
by
\begin{equation}\label{empiricaldedf.eq}
\mu_u(A)=\frac{1}{N}\sum_{i=1}^N\del_{u_i}(A)
\end{equation}
\end{defi}

\noindent where $\delta_u(A)$ is the Dirac mass on
the set $A$, where $\delta_u(A)=1$ if $u$ belongs
to $A$ and $0$ otherwise.\\

Using this formalism provides an useful way
to perform an average over a probability distribution
on the trajectories $u$. For example, the average $\frac{1}{N}\sum_{j=1}^N g[u_j(t)]$,
where $g$ is some function, writes\footnote{Note that the integration variable $\eta$ corresponds to a \textit{trajectory} of
the network, and that $\eta(t)$ corresponds to the \textit{state of the network} at time $t$
(i.e. $\eta_i(t)$ is the state of neuron $i$ at time $t$).} $\int g(\eta(t)) d\mu_u(\eta)$.
 More generally, assume that
we are given a probability distribution $\mu$
on the space of trajectories $\F$. Then,
one can perform 
a generic
construction of a Gaussian probability on $\F$.

\begin{defi}\label{gaussmu.def}
For any $\mu\in\Pp(\F)$ the Gaussian probability distribution $g_\mu$ on
$R^T$, with moments $m_\mu$ and $c_\mu$,  is defined by :
\begin{equation}\label{gaussmfe.eq}\left\{\begin{array}{l}
m_\mu(t+1)=\oups\int f[\eta(t)]d\mu_u(\eta)
\\
c_\mu(s+1,t+1)]=\ups\int f[\eta(s)]f[\eta(t)]d\mu_u(\eta)
\end{array}\right.\end{equation}
\end{defi}
Then, it is easy to reformulate the previous computation as:

\begin{prop}
The common probability distribution of the individual synaptic potential
trajectories $v_i(.,u)$ is the normal distribution $g_{\mu_u}$ where
$\mu_u$ is the empirical distribution of the network potential
trajectory $u$.
\end{prop}

This framework is useful to compute the large-size limit of the
common probability distribution of the potential trajectories.

\subsection{Dynamical models of the membrane potential}\label{SDynpot}

We shall now detail the updating rule of the membrane potential.
Various neural dynamics have been detailed in paper I. 
We focus here on  Analog Formal (AF), Binary Formal (BF),
and Integrate and Fire neuron (IF).

In any case, the network is initialized with independent
identically distributed membrane potential according to a
probability distribution $m_0\in\Pp(\RR)$.
It is useful, on technical grounds, to add to each neuron a small amount of noise.
Thus we introduce for each neuron $i$, a sequence
$(w_i)(t))_{t\in\{1,...,T\}}$ of i.i.d. centered Gaussian variables
of variance $\sig^2$. This sequence is called the \textit{synaptic
noise} of neuron $i$.
The synaptic noise plays an important part in the mathematical proof
but the order parameter $\sig$ is as small as necessary. So this
model is not very restrictive. The synaptic noise is
added to the synaptic potential.  On biological grounds it may account
for different effects such as the diffusion of neurotransmitters involved in the synaptic
transmission, the degrees of freedom neglected by the model,
external perturbations, etc ... Though it is not evident that the ``real noise''
is Brownian, using this kind of perturbations has the advantage
of providing a tractable model where  standard theorems in the theory
of stochastic processes or methods in non equilibrium statistical physics (e.g. Fokker-Planck
equations) can be applied. In some papers, the synaptic
noise is called \textit{thermal noise or annealed noise} by comparison with the random
variables $\J=(J_{ij})$, which are called \textit{quenched
variables} as they are fixed, once for all, and do not change with
time (we do not consider learning in this paper).\\

The  formal neuron updates its membrane potential according
to
\begin{equation}\label{ABFdyn.eq}
u_i(t+1)=v_i(t+1)+w_i(t+1)-\ta
\end{equation}
IF neuron takes into account its present membrane potential while
updating. Its evolution equation is
\begin{equation}\label{IFdyn.eq}
u_i(t+1)=\vfi[u_i(t)+\ta)]+v_i(t+1)+w_i(t+1)-\ta
\end{equation}
where \begin{itemize}
\item $\vfi$ is defined by
\begin{equation}\label{IFdif.eq}
\vfi(u)=\left\{\begin{array}{l}
\ga u \mbox{ if }\frac{\vta}{\ga}<u<\ta \\
\vta\mbox{ else }
\end{array}\right.\end{equation}
\item $\ga\in]0,1[$ is the \textit{leak} (damping coefficient).
\item $\vta$ is the reset potential and $\vta<0<\ta$
\end{itemize}

The following table summarizes the main properties of the three
models we investigate:

\begin{center}
$$
\begin{array}{|c|c|c|}
\hline
\mbox{Transfer function} & \mbox{Heaviside}&\mbox{sigmoidal}
\\ \hline
\mbox{Formal model} & \mbox{BF}&\mbox{AF}
\\ \hline
\mbox{Integrate and Fire} & \mbox{IF}&
\\ \hline
\end{array}$$
\end{center}

\smallskip
Assume for a moment that we remove the neural coupling, then the individual neuron
state trajectories are independent, identically distributed, random
vectors in $\F$ (whose randomness is induced by the Brownian noise).
 The corresponding dynamics is called the \textit{free dynamics}. 
Let us denote by $P$ the common distribution of the neurons trajectory in the
uncoupled case. The probability distribution of the corresponding
neural network trajectory is therefore $P^{\otimes N}$.

In the case of formal neurons, the free dynamics equation is
\begin{equation}\label{ABFfree.eq}
u_i(0)\sim m_0,\ \ u_i(t+1)=w_i(t+1)-\ta
\end{equation}
\nid where $\sim$ means ``distributed according to''.
So
$P=m_0\otimes\N(-\ta,\sig^2)^{\otimes T}$.
In the case of IF neurons $P$ is not explicit. It is the image of
$m_0\otimes\N(-\ta,\sig^2)^{\otimes T}$ by the diffusive
\footnote{
This is indeed a discrete time stochastic difference equations with
drift $\vfi[u_i(t)+\ta]-\ta$ and diffusion $w_i(t+1)$. Incidentally,
we shall come back to stochastic differential equations for neural
networks in the section \ref{MFOscIF} and we shall consider the related 
Fokker-Planck equation.}
 dynamics
\begin{equation}\label{IFfree.eq}
u_i(0)\sim m_0,\ \ u_i(t+1)=\vfi[u_i(t)+\ta)]+w_i(t+1)-\ta
\end{equation}

When coupling the neurons, the trajectory of the
system of neurons is still a random vector.
Its probability distribution, denoted
by $Q_N$, has a density with respect to $P^{\otimes N}$
that can be explicitly computed. This is the main
topic of the next subsection.

\subsection{Computation of the probability distribution of network trajectories.}\label{SDistNN}

This section is devoted to the computation of the probability distribution
$Q_N$. The result shows that the density of $Q_N$ with
respect to the free dynamics probability $P^{\otimes N}$ depends on the
trajectory variable $u$ only through the empirical measure
$\mu_u$. To achieve this computation we shall use a key result of
stochastic process theory, the Girsanov theorem  \cite{Girsanov,Skorhokod}, which gives the
density of the new distribution of a diffusion when the drift is changed.
Actually, since the time set is finite, the version of Girsanov
theorem that we use is different from the original one \cite{Skorhokod} and may be recovered by elementary
Gaussian computation. Its derivation is detailed in the appendix in
theorem \ref{Girsanovdis.theo}. A similar result may be obtained for
continuous time dynamics using the classical Girsanov theorem (see
\cite{Benarous95}).

Let us state the finite-time Girsanov theorem

\begin{theo}\label{Girsanovdisbis.theo}

Let $m_0$ a probability measure on $\RR^d$ and let $\N(\al,K)$ be
a Gaussian regular probability on $\RR^d$ with mean $\al$ and covariance
matrix $K$. Let $T$ be a positive
integer and $\E_T=(\RR^d)^{\{0,...,T\}}$ be the space of finite time
trajectories in $\RR^d$. Let $w$ be a Gaussian random vector in
$\E_T$ with distribution $m_0\otimes\N(\al,K)^T$. Let
$\phi$ and $\psi$ be two measurable applications of $\RR^d$ into
$\RR^d$. Then we define the random vectors $x$ and $y$ in $\E$ by:

\begin{equation}\left\{\begin{array}{l}
x_0=w_0\\x(t+1)=\phi[x(t)]+w(t+1)
\end{array}\right.\end{equation}

\begin{equation}\left\{\begin{array}{l}
y_0=w_0\\y(t+1)=\psi[y(t)]+w(t+1)
\end{array}\right.\end{equation}

\smallskip

Let $P$ and $Q$ be the respective probability distributions on $\E$ of $x$
and $y$, then we have:

\bea
\label{GirsanovDiff.eq}
&&\frac{dQ}{dP}(\eta)=\\
&&\exp\sum_{t=0}^{T-1} 
\left\{
\begin{array}{l}
-\frac{1}{2}\{\psi[(\eta(t)]-\phi[(\eta(t)]\}^t
K^{-1}\{\psi[(\eta(t)]-\phi[(\eta(t)]\}\\
+\{\psi[(\eta(t)]-\phi[(\eta(t)]\}^tK^{-1}
\{\eta(t+1)-\al-\phi[\eta(t)]\}
\nonumber
\end{array}
\right\}
\eea

\nid namely  $Q$ is \underline{absolutely continuous} with respect to $P$.
\end{theo}

We shall use this theorem to prove the following:

\begin{theo}\label{densityN.theo}
The density of the distribution of the network membrane potential $Q_N$ with
respect to $P^{\otimes N}$ is given by

\begin{equation}
\frac{dQ_N}{dP^{\otimes N}}(u)=\exp N\Ga(\mu_u)
\end{equation}

\nid where the functional $\Ga$ is defined on $\Pp(\F)$
by:

\beq
\Ga(\mu)=\int\log\left\{\int\exp\frac{1}{\sig^2}
\sum_{t=0}^{T-1}\left[-\frac{1}{2}\xi(t+1)^2
+\Phi_{t+1}(\eta)\xi(t+1)\right]
dg_\mu(\xi)\right\}d\mu(\eta)\nonumber
\eeq

\nid with:

\begin{itemize}
\item for AF and BF models: $\Phi_{t+1}(\eta)=\eta(t+1)+\ta$
\item IF model: $\Phi_{t+1}(\eta)=\eta(t+1)+\ta-\vfi[\eta(t)+\ta]$
\end{itemize}
\end{theo}

\begin{rem}
Let us recall that the Gaussian measure $g_\mu$ has been defined
previously (Definition \ref{gaussmu.def})
\end{rem}

\smallskip

\prtheo
Call $Q_N(\J)$ the conditional distribution of the network state
trajectory given $\J$, the matrix of synaptic weights. We shall
apply the finite-time Girsanov theorem \ref{Girsanovdisbis.theo} to
express $\frac{dQ_N(\J)}{dP\otimes N}$. To apply the theorem we
notice that
\begin{itemize}
\item The difference of the two drift terms
$\psi[\eta(t)-\phi[\eta(t)]$ of the theorem is here the synaptic
potentials $(v_i(t))$. The synaptic potentials $v_i$ are functions
of the $u_i$'s according to (\ref{synapticdyn.eq})
\[v_i(\J,u)(t+1)=\sum_{j=1}^N J_{ij}f[u_j(t)]\]

\item The expression of the synaptic noise $(w_i(t+1)$, as a
function of the state trajectory $u_i$ in the free dynamics, is
given by $\Phi_{t+1}(u_i)$.  The explicit form of $\Phi$ depends on the neuron model
(formal or IF).
\end{itemize}
We have so

\begin{equation}\label{densitydyn.eq1}
\frac{dQ_N(\J)}{dP^{\otimes N}}(u)=\exp\frac{1}{\sig^2}
\sum_{t=0}^{T-1}\sum_{i=1}^N\left[
-\frac{1}{2} v_i(\J,u)(t+1)^2
+v_i(\J,u)(t+1)\Phi_{t+1}(u_i)\right]
\end{equation}

\begin{equation}\label{densitydyn.eq2}
\frac{dQ_N(\J)}{dP^{\otimes N}}(u)=
\prod_{i=1}^N\exp\frac{1}{\sig^2}\sum_{t=0}^{T-1}\left[
-\frac{1}{2} v_i(\J,u)(t+1)^2
+v_i(\J,u)(t+1)\Phi_{t+1}(u_i)\right]
\end{equation}

\smallskip

We have thus obtained the probability distribution
of the membrane potentials in the coupled neurons
models, but \textit{for a fixed realization of
the $J_{ij}$'s} (conditional probability).
 
Let us now consider the probability of the quenched variables
$\J=(J_{ij})$. We observed previously when we introduced the
synaptic potential model that under the configuration distribution
of $\J$, the random vectors $v_i(\J,u)$ are independent identically
distributed according to the normal distribution $g_{\mu_u}$
To compute the density of $Q_N$ with respect to $P^{\otimes N}$ one
has thus to average the conditional density
$\frac{dQ_N(\J)}{dP^{\otimes N}}$ over the
distribution of $\J$. 
\textit{Since the $J_{ij}$'s are independent, the integration separates into products} and
one gets from (\ref{densitydyn.eq2}) the following

\[
\frac{dQ_N}{dP^{\otimes N}}(u)=
\exp\sum_{i=1}^N\log\int\exp \frac{1}{\sig^2}\sum_{t=0}^{T-1}
\left[-\frac{1}{2}\xi(t+1)^2+\Phi_{t+1}(u_i)\xi(t+1)\right]
dg_{\mu_u}(\xi)
\]

The sum over $i$ is equivalent to an integration over the empirical
measure $\mu_u$ (\ref{empiricaldedf.eq}), so we have
\[
\frac{dQ_N}{dP^{\otimes N}}(u)=
\exp N\int
\log\left\{\int\exp\frac{1}{\sig^2}\sum_{t=0}^{T-1}
\left[-\frac{1}{2}\xi(t+1)^2+\Phi_{t+1}(\eta)\xi(t+1)\right]
dg_{\mu_u}(\xi)\right\}d\mu_u(\eta)
\]
\cqfd

\textbf{Remark.} These equations reminds the generating functional approach
derived e.g. by Sompolinsky \& al. \cite{Sompo88,Cri90} or Molgedey \& al \cite{Molgedey92}
allowing to compute the moments of the $u_i(t)$'s. However, the
present approach provides a stronger result. While the generating
functional method   deals with weak convergence (convergence of generating
function) the method developed here allows to obtain
directly the probability distribution of the $u_i(t)$'s.
Moreover, by using large deviations techniques one is able
to establish almost-sure convergence results (valid for
only one typical sample).  \\

Let us now state an important corollary of this theorem.
 
\begin{coro}\label{empiricdensity.coro}
The  empirical measure $\mu_u$ is a random measure governed by $Q_N$.
It has a density  with
respect to the distribution of the empirical measure of the free model, (that
is governed by $P^{\otimes N}$),  given by:
$$\mu\in\Pp(\F) \ra \exp N\Ga(\mu)$$
\end{coro}

\subsection{Summary of notations}\label{SSumNot.tex}
Let us recall the notations of this section. They will be
extensively used in the following sections:

\vspace{10mm}

$\begin{array}{|c|l|}\hline
\mbox{Notation}& \mbox{Interpretation}
\\ \hline
i\in\{1,...,N\} & $individual neuron in  a $N$ neuron
population$
\\
t\in\{0,...,T\} & $time label of the discrete time dynamics
at horizon $T
\\
u_i(t)&$membrane potential of neuron $i$ at time $t
\\
x_i(t)&$activation state of neuron $i$ at time $t
\\
v_i(t)&$synaptic potential of neuron $i$ at time $t
\\
w_i(t)&$synaptic summation noise of neuron $i$ at time $t
\\
u_i\in\F&$membrane potential trajectory of neuron $i$ from time $0$
to time $T
\\
u\in\F^N&$network membrane potential\underline{s} trajectory (\underline{network} trajectory)
from time $0$ to time $T
\\
x_i\in\E&$ activation state trajectory of neuron $i$ from time
$0$ to time $T
\\
x\in\F^N&$network activation state trajectory$
$ from time $0$ to time $T
\\
\ta&$common firing threshold of individual neurons$
\\
\sig&$common standard deviation of the synaptic noise$
\\
\la&$leak current factor for Integrate and fire (IF) neuron
model$
\\
f&$neuron transfer function converting membrane potential into
activation state$
\\
J_{ij} &$synaptic weight from neuron $j$ to neuron $i$ (real
random variable)$
\\
\J=(J_{ij})&$synaptic weight matrix (random $N\times N$ matrix)$
\\
\frac{\ol\ups}{N}& $expectation of synaptic weights
$J_{ij}
\\
\frac{\ups}{N}& $variance of synaptic weights
$J_{ij}
\\
\mu\in\Pp(\F)&$generic probability distribution of individual
membrane potential trajectory$
\\
\eta\in\F&$random vector which takes its values in $\F$ under
probability distribution $\mu
\\
P\in\Pp(\F)&$probability distribution of individual
membrane potential trajectory for free dynamics$
\\
g_\mu\in\Pp(\F)&$synaptic potential distribution obtained
from $\mu\in\Pp(F)$ through central limit approximation$
\\
Q_N\in\Pp(\F^N)& $probability distribution of network membrane potential
trajectory $u
\\\hline
\end{array}
$

\section{The mean-field dynamics }\label{DMFT}

\subsection{Introduction to mean-field theory} \label{IMFT}

The aim of this section is to describe  the evolution of a typical neuron
in the limit of large size
networks.  This is done by summarizing, in a
single term, the effect of the interactions of this neuron with the
other neurons of the network.  A mean-field theory provides evolution equations
of the type of free dynamics equation, that are involving single
neuron dynamics, but where an \textit{effective interaction term} remains. 
This term, or
"mean-field", is properly the average effect of all the interaction
of other neurons with the neuron of interest. So the mean field dynamics is intermediate between the
\textit{detailed} dynamics, which takes into account all the
detailed interactions between neurons, and the \textit{free} dynamics, which
neglects all  interactions.

 To derive mean-field equations in a direct way, one can
replace $v_i(t)$ by an approximation which depends
only on the statistical distribution of the $u_j(t)$'s.
This  approximation  takes advantage of the
large number of synapses $J_{ij}$ to postulate the vanishing of individual
correlations between neurons or between neurons and configuration
variables. This is the hypothesis of "local chaos" of Amari
(\cite{Amari72},\cite{Amari77}), or of "vanishing correlations"
which is usually invoked to support mean-field equations.
In the present context, it can be stated as follows.\\

\textit{ In the large size limit, the $u_j$'s are asymptotically
independent, they are also independent from the configuration
parameters }\\

\nid This approach is
very similar to the  Boltzmann's
"molecular chaos hypothesis"\footnote{The word ``chaos''
is somehow confusing here, especially because we also dealt with deterministic
chaos in the paper I. Actually, ``deterministic chaos''
and the related exponential correlation decay can be invoked, 
in statistical physics, to
 obtain (deterministic) equations for the mean value and Gaussian
fluctuations of  relevant observables. In the present case 
the basic reason that makes the mean-field approach ``works''
is however different. This is the mere fact that the model
is fully connected and that the $J_{ij}$'s are independent and vanishing in 
the limit $N \to \infty$. 
This is standard result in statistical physics
models such as the Curie-Weiss model but obtaining this for the trajectories
of a dynamical
model with quenched disorder requires more elaborated techniques.} introduced by 
Boltzmann (1872), tailored to neural networks dynamics. \\

From the central limit theorem we are then allowed to state that
the random variable $\zeta(t+1)=\sum_{j=1}^N J_{ij}f(u_j(t))$ is  a large sum of approximatively independent identically
distributed variable, and thus that it has approximatively a
Gaussian distribution. Thus, we just have  to derive its first and second order
moment from the common probability distribution of $u_i=(u_i(t))$ to know
completely the distribution of $\zeta$. Henceforth, from a probability distribution on $\F$ which is supposed to be the
common probability distribution of the $u_i$'s, we are able to derive the distribution
of $\zeta$ and then the distribution of the resulting potential trajectory
and  the state trajectory of a generic vector of the network.

The assumption on which the mean-field approximation is based may look entirely wrong at first glance.
However, in the present context it gives exactly the same results
as more elaborated methods such as the generating functional method, or
the large deviations approach developed below. 
Moreover, it is supported by the ``propagation of chaos'' result
proved in section \ref{ResMFT}.
Note however that in models
with correlated interactions $J_{ij}$ (such as spin-glasses,
 where $J_{ij}=J_{ji}$) the ``local chaos'' hypothesis leads
to wrong results (at low temperature) while generating functional
methods  \cite{Zippelius82,Crisanti87a,Crisanti87b} and large deviations techniques \cite{Benarous95} can still be used.

\subsection{Mean-field propagation operator and mean-field equation}\label{MFE}


We now define an evolution operator $L$, on the set
$\Pp(F)$ of  probability distributions on $\F$, that we call the
\textit{mean-field propagation} operator (or mean-field propagator).
Let $\mu\in\Pp(\F)$ be a probability measure on $\F$. Let us
compute the moments of
$$\forall t\in\{0,1,...,T-1\},
\zeta(t+1)=\sum_{j=1}^N J_{ij}f[u_j(t)]$$
where the $u_j$'s are independent identically
distributed random vectors with probability distribution $\mu$.
They are also
independent from the configuration parameters $J_{ij}$.

Since $E[J_{ij}]=\frac{\oups}{N}$ and
$\textrm{Var}[J_{ij}]=\frac{\ups}{N}$, we have
\begin{equation}\label{gaussmfe.eqbis}\left\{\begin{array}{l}
E[\zeta(t+1)]=\oups\int_\F f[\eta(t)]d\mu(\eta)
\\
\Cov[\zeta(s+1),\zeta(t+1)]=\ups\int_\E
f[\eta(s)]f[\eta(t)]d\mu(\eta)
\end{array}\right.\end{equation}
Notice that the expression of the covariance is asymptotic since
the sum of squares of expectation of the synaptic weights may be
neglected. So $\zeta$ is a Gaussian random vector in $\F$ with
probability distribution $g_\mu$ (see definition \ref{gaussmu.def}).

\begin{defi}\label{L.defi}
Let $\mu$ a probability distribution on $\F$ such that the distribution of the
first component is $m_0$. Let $u,w,v$ be three independent
random vectors with the following distributions
\begin{itemize}
\item the distribution of $u$ is $\mu$,
\item the distribution of $w$ is $\N(0,\sig^2 I_T)$,
\item the distribution of $v$ is $g_\mu$
\end{itemize}
Then $L(\mu)$ is the probability distribution on $\F$ of the random vector
$\vta$ which is defined by
\begin{equation}\label{ABFMFE.eq}\left\{\begin{array}{l}
\vta(0)=u(0)
\\
\vta(t+1)=v(t+1)+w(t+1)-\ta
\end{array}\right.\end{equation}
for the formal neuron models (BF and AF), and by
\begin{equation}\label{IFMFE.eq}\left\{\begin{array}{l}
\vta(0)=u(0)
\\
\vta(t+1)=\vfi[u(t)+\ta)]+v(t+1)+w(t+1)-\ta
\end{array}\right.\end{equation}
for the IF neuron model.
The operator $L$ which is defined on $\Pp(\F)$ is called the
\underline{mean-field propagation operator}.
\end{defi}

\begin{defi}
The following equation on $\mu\in\Pp(\F)$
\begin{equation}\label{mfe.eq}
L(\mu)=\mu
\end{equation}
is called the \textbf{mean-field equation (MFE)}
\end{defi}

\begin{rem}\label{mfe.rem}
The mean-field equation is the achievement of mean-field approach.
To determine the distribution of an individual trajectory, we suppose that
this distribution governs the interaction of all the units onto the selected
one. The resulting distribution of the selected unit has to be the same distribution
than the generic distribution. This is summarized in the mean-field equation
\[ L(\mu)=\mu\]
\end{rem}

Equations (\ref{ABFMFE.eq}) (resp. (\ref{IFMFE.eq}) for the IF model)
with the specification of the probability distributions define the
\textit{mean-field
dynamics}. Actually, the distribution $L(\mu)$ is just the convolution of the
probability distributions $P$ and the Gaussian distribution $g_\mu$.  More precisely,
if we apply the discrete time Girsanov theorem
\ref{Girsanovdis.theo} of the appendix, we have:

\begin{theo}\label{density1.theo}
$L(\mu)$ is absolutely continuous with respect to $P$ and its density
is given by
\begin{equation}\label{MFEProp}
\frac{dL(\mu)}{dP}(\eta)=
\int\exp\frac{1}{\sig^2}\sum_{t=0}^{T-1}
\left[-\frac{1}{2}\xi(t+1)^2+\Phi_{t+1}(\eta)\xi(t+1)\right]
dg_\mu(\xi)
\end{equation}
\end{theo}
\pr
The proof is essentially a simplified version of the application of
the finite-time Girsanov theorem which was used to prove
theorem (\ref{densityN.theo}). The conditioning is done here with
respect to $v$ which is the difference between the drift terms
of the free dynamics and of the mean-field dynamics.
  \cqfd

\begin{rem}\label{Gammamfe.rem}
We have to notice for further use that
\begin{equation}
\Ga(\mu)=\int\log\frac{dL(\mu)}{dP}(\eta)d\mu(\eta)
\end{equation}
\end{rem}

In all the  cases, for $0<t<T$ the projection of the distributions
$\Ga(\mu)$ and $L(\mu)$ on the $t+1$ first time steps just depends on
the projection of $\mu$ on the $t$ first instants. Since the
projection of $\mu$ on the initial instant is always $m_0$, the
projection of $L(\mu)$ on the two first instants $\{0,1\}$ depend
only on $m_0$ and similarly, the projection of $L^t(\mu)$ on the
$t+1$ first instants $\{0,1,...,t\}$ depends only on $m_0$.
Eventually
$\mu_T=L^T(\mu)=L^T(P)$ depends only on $m_0$ and it is the only
fixed point of the mean-field propagation operator $L$.

So we have shown the following

\begin{theo}\label{mfe.theo}
The probability measure $\mu_T$=$L^T(P)$ is the only solution of
the mean-field equation with initial condition $m_0$.
\end{theo}

\subsection{Large Deviation Principle  for RRNN  mean-field theory}\label{LDP}
In this section, we fully use the computation results of the
previous section to show the rigorous foundations of mean-field
theory for RRNN. The approach is the following:

\begin{itemize}

\item[(a)] The empirical measure $\mu_u$ of the network dynamics
satisfies a \textit{large deviation principle (LDP)} under
$P^{\otimes N}$ with a good rate function $\mu\in\Pp(\F)\ra
I(\mu,P)\in\RR^+$, the relative entropy between $\mu$ and $P$. 
Actually, when the size of the network tends to
infinity, the empirical measure converges in distribution exponentially fast
towards
$P$. The definition of LDP and  its consequences are outlined in the
appendix in definition \ref{LDP}. Sanov theorem is stated in
appendix, theorem \ref{Sanov.theo}.

\item[(b)] According to corollary \ref{empiricdensity.coro}, the
density of the new distribution of $\mu_u$ with respect to the original distribution
when we switch from $P^{\otimes N}$, that governs the free dynamics,
  to $Q_N$, that governs the RRNN dynamics is $\exp N\Ga(\mu)$.

\item[(c)] Combining  (a) and (b), one obtains that under $Q_N$, the
sequence $\mu_u$ satisfies a LDP with the good rate function
\begin{equation}\label{mfetaux.eq}
  H(\mu)= I(\mu,P)-\Ga(\mu)
\end{equation}
This kind of result is used in statistical physics under the name
of \textit{Gibbs variational principle} \cite{Ellis85}. The
functional $H$ is called, in statistical physics, a \textit{thermodynamic potential }
(e.g. free energy or Gibbs potential). Notice that the
classical statistical mechanics framework is relative to equilibrium probability
distributions on the space of microscopic states.
 It is applied here to trajectories. For that reason,
this approach is called the \textit{dynamic mean-field theory}
\cite{Sompo88}. It is quite technical to support it
rigorously. One has to show that $H$ is lower semi-continuous and
is a good rate function (see Varadhan's theorem \ref{Varadhan.theo}
of the appendix). This kind of proof is rather technical. 
To reduce the size of the paper we admit the following result
(see
\cite{Benarous95} for a general approach and \cite{Moynot02} for
the proof for AFRRNN model)

\begin{theo}
Under the respective distributions $Q_N$ the family of empirical measures
$(\mu_N)$ of $\mc{P}(\F)$ satisfies a full large deviation principle
with a good rate function H given by (\ref{mfetaux.eq}).
\end{theo}

\item[(d)] It is clear from remark \ref{Gammamfe.rem} that
$H(\mu_T)=0$ where $\mu_T$ is is the unique solution of MFE with
initial condition $m_0$, so it is the fixed point of L. Thus $\mu_T$
is a minimum of $H$.\\

The basic computation is the following: first we apply the
definition \ref{crossentropy.defi} of the relative entropy that is
given in  the appendix

$$I(\mu_T,P)=\int \log\frac{d\mu_T}{dP}(\eta)d\mu_T(\eta)$$
Since $\mu_T$ is the solution of MFE, we have
$$
\frac{d\mu_T}{dP}(\eta)=\frac{dL(\mu_T)}{dP}(\eta)
$$
then we apply the previous remark \ref{Gammamfe.rem} which states
$$
\Ga(\mu_T)=\int\log\frac{dL(\mu_T)}{dP}(\eta)d\mu_T(\eta)
$$
to check
$$ I(\mu_T,P)=\Ga(\mu_T)\Rightarrow H(\mu_T)=0$$

\item[(e)]
To obtain the exponential convergence of the sequence of empirical
measures $\mu_u$ under $Q_N$ when $N\ra\infty$, one has eventually
to show that $H(\mu)=0\Rightarrow \mu=\mu_T $. This point is
technical too. It is proved in a similar still more general
framework (continuous time) in \cite{Benarous95} using a Taylor
expansion. The same method is and applied to show the uniqueness for
AFRRNN model in \cite{Moynot02}.
\end{itemize}

Thus, we have the main result of that section:

\begin{theo}\label{RRNNMFE.theo}
When the size $N$ of the network goes to infinity, the sequence of
empirical measures $(\mu_u)$ converges in probability exponentially
fast towards $\mu_T$ which is the unique solution of the mean-field
equation $\ L(\mu)=\mu$
\end{theo}

\subsection{Main results of RRNN  mean-field theory}\label{ResMFT}

First notice that theorem \ref{RRNNMFE.theo} may be extended to
RRNN with fast decreasing connection weights distribution. More
precisely, assume that the common distribution $\nu_N$ of the
connexion weights satisfies the following:

\begin{hypo}[H]\label{H.hypo}
For all $N$, the common probability law $\nu_N$ of the connexion
weights satisfies
$$\begin{array}{ll}
(i)&\int w d\nu_N(w)=\frac{\oups}{N}
\\
(ii)&\int w^2 d\nu_N(w)=\frac{\ups}{N}+\frac{\oups^2}{N^2}
\\
(iii)&\exists a>0,\exists D>0 \textrm{ such that }
\int \exp(aNw^2) d\nu_N(w)\leq D
\end{array}$$
then the family $(\nu_N)$ is said to satisfy hypothesis (H).
\end{hypo}

Then, from theorem \ref{RRNNMFE.theo} two important results can be
deduced rigorously. The first one is a "propagation of chaos"
result which supports the basic intuition of mean field theory about
the asymptotic independence of finite subsets of individuals when the
population size grows to infinity.

\begin{theo}\label{RRNNKOProp.theo}
Let $k$ be a positive integer and $(f_i)_{i\in\{1,...,k\}}$ be $k$
continuous bounded functions on $\F$, when the size
$N$ of the network goes to infinity, then
\[
\int\prod_{i=1}^{k}f_i(u_i)dQ_N(u)\ra
\prod_{i=1}^{k} \int f_i(\eta)d\mu_0(\eta)
\]
\end{theo}
\pr
The idea of the proof is due to Snitzman \cite{Sznitman84}.

First, a straightforward consequence of theorem \ref{RRNNMFE.theo}
is that when we apply the sequence of random measures $(\mu_N)$ to
the test function $F$ on $\Pp(\F)$ defined by
$F(\mu)=\prod_{i=1}^{k}\int f_i(u_i)d\mu(u)$ we get the convergence
of
\[\lim_{N\ra\infty}\int \prod_{i=1}^{k}
\frac{1}{N}\left[\sum_{j=1}^N f_i(u_j)\right]dQ_N(u)
=\prod_{i=1}^{k}\int f_i(\eta)d\mu_0(\eta)
\]
Thus it remains to compare
$\int \prod_{i=1}^{k}
\frac{1}{N}\left[\sum_{j=1}^N f_i(u_j)\right]dQ_N(u)$
and $\int\prod_{i=1}^{k}f_i(u_i)dQ_N(u)$
From the symmetry property of $Q_N$, it is clear that for any subset
$\{j_1,...,j_k\}$ of $k$ neurons among N, we have
\[
\int\prod_{i=1}^{k}f_i(u_{j_i})dQ_N(u)
=\int\prod_{i=1}^{k}f_i(u_i)dQ_N(u)
\]
If we develop $\int \prod_{i=1}^{k}\frac{1}{N}\left[\sum_{j=1}^N
f_i(u_j)\right]dQ_N(u)$, we get

\begin{equation}\label{formul1}
\int \prod_{i=1}^{k}\frac{1}{N}\left[\sum_{j=1}^N
f_i(u_j)\right]dQ_N(u)=
\frac{1}{N^k}\sum_{\{j_1,...,j_k\}}\int
\prod_{i=1}^{k}f_i(u_{j_i})dQ_N(u)
\end{equation}

The average sum in (\ref{formul1}) is here over all applications of
$\{1,...,k\}$ in $\{1,...,N\}$. And the equality is proved if we
replace it by the average over all injections of $\{1,...,k\}$ in
$\{1,...,N\}$, since the terms are all equal for injections. But
when $N$ goes to infinity the proportion of injections which is
$\frac{N!}{(N-k)!N^k}$ goes to 1 and thus the contributions of
repeated k-uple is negligible when $N$ is large. Therefore
\[
\lim_{N\ra\infty}\left[ \int\prod_{i=1}^{k}
\frac{1}{N}\left[\sum_{j=1}^N f_i(u_j)\right]dQ_N(u)-
\int\prod_{i=1}^{k}f_i(u_i)dQ_N(u)\right]
=0
\]\cqfd\\

Still, this propagation of chaos result is valid when the
expectation of the test function is taken with respect to the
connection distribution. Thus, it doesn't say anything precise about the
observation relative to a \textit{single} large-sized network.

Actually, since exponentially fast convergence in probability
implies almost-sure convergence form Borel-Cantelli lemma, we  are
able to infer the following statement from theorem
\ref{RRNNMFE.theo}. Recall that we note (as in the proof of theorem
\ref{densityN.theo}) $Q_N(\J)$ the conditional distribution of the network
state trajectory given $\J$ the system of synaptic weights and
we define $\mu_N(u)=\frac{1}{N}\sum_{i=1}^N\del_{u_i}$ for the
empirical measure on $\F$ which is associated to a network
trajectory $u\in\F^N$.

\begin{theo}\label{StrLLN.theo}
Let $F$ be a bounded continuous functional on $\Pp(\F)$, we
have almost surely in $\J$
\[
\lim_{N\ra\infty}\int F[\mu_N(u)]dQ_N(\J)(u)=F(\mu_T)
\]
\end{theo}
Note that we cannot use this theorem to infer a "quenched" propagation
of chaos result similar to theorem \ref{RRNNKOProp.theo}, which
was an ``annealed'' propagation of chaos result (i.e. averaged over the
connection weight distribution). This is not possible because, for a
given network configuration $\J$, $Q_N(\J)$ is no more symmetric
with respect to the individual neurons. Nevertheless, we obtain the
following crucial result, applying theorem \ref{StrLLN.theo} to the
case where $F$ is the linear form $F(\mu)=\int fd\mu$

\begin{theo}\label{StrLLN2.theo}
Let $f$ be a bounded continuous function on $\F$, we
have almost surely in $\J$
\[
\lim_{N\ra\infty}\frac{1}{N}\sum_{i=1}^N
\int f(u_i) dQ_N(\J)(u)=\int f(\eta)d\mu_T(\eta)
\]
\end{theo}

The consequences of these results are developed in the next section. 

\section{Mean-field dynamics for analog networks } \label{MFAnalog}

We are interested in the stationary dynamics of large
random recurrent neural networks. Moreover since we want to study
the meaning of oscillations and of chaos, the regime of low noise
is specially interesting since the oscillations are practically
cancelled if the noise is too loud. For these reasons, we cannot be
practically satisfied by  obtaining the limit $\mu_T$ of the
empirical measures. So we shall extract from $\mu_T$ dynamical
informations on the asymptotics of the network trajectories. Notice
that the distribution of the connexion weight distribution is not
necessarily Gaussian as long as it satisfies hypothesis
(H:\ref{H.hypo}).

\subsection{Mean-field dynamics of homogeneous networks} \label{MFDHN}
\subsubsection{General mean-field equations for moments} \label{MFMoments}
Recall that in section 2 of this chapter (definition
\ref{gaussmu.def}) we defined for any probability measure
$\mu\in\Pp(\F)$ the two first moments of $\mu$, $m_\mu$ and $c_\mu$.
Let us recall these notations:
$$\left\{\begin{array}{l}
m_\mu(t+1)=\oups\int f[\eta(t)]d\mu(\eta)
\\
c_\mu(s+1,t+1)=\ups\int f[\eta(s)]f[\eta(t)]d\mu(\eta)
\\
q_\mu(t+1)=c_\mu(t+1,t+1)
\end{array}\right.$$
where $f$ is the sigmoid function
$f(x)=\displaystyle{\frac{e^x}{1+e^x}}$

In this section, in order to alleviate notations, we note $m,c,q$
instead of $m_{\mu_T},c_{\mu_T},q_{\mu_T}$ where $\mu_T$ is the
asymptotic probability that was shown to be a fixed point of the
mean-field evolution operator $L$ in last section. By expressing
that $\mu_T$ is a fixed point of $L$, we shall produce some
evolution autonomous dynamics on the moments $m,c,q$.

More precisely we have from the definition of $L$ (see definition
\ref{L.defi} in section 3) that the law of $\eta(t)$ under $\mu_T$
is a Gaussian law of mean $m(t)-\ta$ and of variance
$q(t)+\sig^2$ (see equations (\ref{gaussmfe.eqbis}) and
(\ref{ABFMFE.eq})). So we have
\begin{equation}\label{MFmeanstd.eq}\left\{\begin{array}{l}
m(t+1)=\oups\int f[\sqrt{q(t)+\sig^2}\xi+m(t)-\ta]d\ga(\xi)
\\
q(t+1)=\ups\int f[\sqrt{q(t)+\sig^2}\xi+m(t)-\ta]^2d\ga(\xi)
\end{array}\right.\end{equation}
where $\ga$ is the standard Gaussian probability on $\RR$:
$d\ga(\xi)=\frac{1}{\sqrt{2\pi}}\exp\left[-\frac{\xi^2}{2}\right]
d\xi$.

Moreover, the covariance of $(\eta(s),\eta(t))$ under $\mu_T$ is
$c(s,t)$ if $s\neq t$. Thus in this case, considering the standard
integration formula of a 2 dimensional Gaussian vector:

$$\begin{array}{ll}
E[f(X)g(Y)]=\\
\int\int 
f\left(\sqrt{\frac{Var(X)Var(Y)-Cov(X,Y)^2}{Var(Y)}}\xi_1+
\frac{Cov(X,Y)}{\sqrt{Var(Y)}}\xi_2+E(X)\right\}
g[\sqrt{Var(Y)}\xi_2+E(Y)] d\ga(\xi_1)d\ga(\xi_2)
\end{array}$$

\nid  we obtain the following evolution equation for covariance:

\begin{equation}\label{MFcovariance.eq}\begin{array}{ll}
c(s+1,t+1)=\\
\ups\int\int
f\left(\sqrt{\frac{[q(s)+\sig^2][q(t)+\sig^2]-c(s,t)^2}{q(t)+\sig^2}}
\xi_1+\frac{c(s,t)}{\sqrt{q(t)+\sig^2}}\xi_2+m(s)-\ta\right)f[\sqrt{q(t)+\sig^2}\xi_2+m(t)-\ta]
d\ga(\xi_1)d\ga(\xi_2)
\end{array}\end{equation}
The dynamics of the mean-field system
(\ref{MFmeanstd.eq},\ref{MFcovariance.eq}) can be studied as a
function of the parameters:
\begin{itemize}
\item the mean $\oups$ of the connexion weights,
\item the standard deviation $\ups$ of the connexion weights
\item the firing threshold  $\ta$ of neurons.
\end{itemize}
Notice that the time and size limits do not necessarily commute.
Therefore, any result on long time dynamics of the mean-field
system may not be an exact prediction of the large-size limit of
stationary dynamics of random recurrent networks. However, for our
model, extensive numerical simulations have shown
(\cite{Cessac94},\cite{Dauce98} and chapter I) that the time asymptotics of the
mean-field system is informative about moderately large random
recurrent network stationary dynamics (from size of some hundred
neurons).

More precisely, in the low noise limit $(\sig<<1)$, two points of
view are interesting:
\begin{itemize}
\item the ensemble stationary dynamics is given by the study of the
time asymptotics of the dynamical system
\begin{equation}\label{MFmeanstdsig0.eq}\left\{\begin{array}{l}
m(t+1)=\oups\int f[\sqrt{q(t)}\xi+m(t)-\ta]d\ga(\xi)
\\
q(t+1)=\ups\int f[\sqrt{q(t)}\xi+m(t)-\ta]^2d\ga(\xi)
\end{array}\right.\end{equation}
\item the synchronization of the individual neuron trajectories.
Actually, $m(t)$ and $q(t)$ may converge, when $t\ra\infty$,
towards limits $m^*$ and $q^*$ (stable equilibria of
the dynamical system \ref{MFmeanstdsig0.eq})  with a great variety
of dynamical behaviors. Each individual trajectory may converge
to a fixed point and $(m^*,q^*)$ are the statistical moments of
the fixed point empirical distributions. Another case is provided
by individual chaotic oscillations around  $m^*$ where $q^*$
measures the amplitude of the oscillations.
\end{itemize}
The discrimination between these two situations which are very
different from the point of view of neuron dynamics is given by the
study of the mean quadratic distance which will be outlined in the
next paragraph.

\subsubsection{Study of the mean quadratic distance}\label{MeanD}

The concept of mean quadratic distance was introduced by
Derrida and Pommeau in \cite{Derrida86} to study the chaotic
dynamics of extremely diluted large size networks. The method
originates to check the sensitivity of the dynamical system to
initial conditions. The idea is the following: let us consider two
networks trajectories $u^{(1)}$ and $u^{(2)}$ of the same network
configuration which is given by the synaptic weight matrix
$(J_{ij})$. Their mean quadratic distance is defined by
$$ d_{1,2}(t)=\frac{1}{N}\sum_{i=1}^N[u_i^{(1)}(t)-u_i^{(2)}(t)]^2$$
For a given configuration, if the network trajectory converges
towards a stable equilibrium or towards a limit cycle (synchronous
individual trajectories), then the mean quadratic distance between
closely initialized trajectories goes to 0 when times goes to
infinity. On the contrary, when this distance goes far from 0, for
instance converges towards a non zero limit, whatever close the
initial conditions are, the network dynamics present in some sense
"sensitivity to initial conditions" and thus this behavior of the
mean quadratic distance can be considered to be symptomatic of
chaos. We applied this idea in \cite{Cessac95} to characterize
instability of random recurrent neural network.

In the context of large deviation based mean-field theory, the
trajectories $u^{(1)}$ and $u^{(2)}$ are submitted to independent
synaptic noises and the mean quadratic distance is defined by
\begin{equation}\label{mqdistance.eq}
d_{1,2}(t)=\frac{1}{N}\sum_{i=1}^N
\int [u_i^{(1)}(t)-u_i^{(2)}(t)]^2dQ_N^{(1,2)}(u^{(1)},u^{(2)})
\end{equation}
where $Q_N^{(1,2)}$ is the joint probability law on $\F^{2N}$ of the
network trajectories $(u^{(1)},u^{(2)})$ over the time interval
$\{0,...,T\}$. Following the same lines as in last sections, it is
easy to show a large deviation principle for the empirical measure
of the sample $(u_i^{(1)},u_i^{(2)})_{i\in\{1,...,N}$ under
$Q_N^{(1,2)}$ when $N\ra\infty$. Then we get the almost sure
convergence theorem
\[
\lim_{N\ra\infty}\frac{1}{N}\sum_{i=1}^N
\int f_1(u_i^{1}) f_2(u_i^{2}) dQ_N(\J)(u)=
\int f_1(\eta_1)f_2(\eta_2)d\mu_T^{(1,2)}(\eta_1,\eta_2)
\]
where $\mu_T^{(1,2)}$ is the fixed point of the mean-field
evolution operator $L^{(1,2)}$ of the joint trajectories, which is
defined on the probability measure set $\Pp(\F\times\F)$ exactly in
the same way as $L$ was defined previously in definition
\ref{L.defi}.

Then if we define the instantaneous covariance between two
trajectories by:\\

\begin{defi}\label{cov12.def}
The instantaneous cross covariance between the two trajectories
under their joint probability law is defined by
\begin{equation}\label{cov12.eq}
c_{1,2}(t)=\int \eta_1(t)\eta_2(t) d\mu_T^{(1,2)}(\eta_1,\eta_2)
\end{equation}
where $\mu_T^{(1,2)}$ is the fixed point measure of the joint
evolution operator $L^{(1,2)}$ defined from an initial condition
$\mu_{init}^{(1,2)}$,
\end{defi}

\smallskip

\nid then we can follow the argument, which was already used for the
covariance evolution equation (\ref{MFcovariance.eq}). Thus we
obtain the following evolution equation for the instantaneous cross
covariance equation

\begin{equation}\label{MFcov12.eq}
\begin{array}{ll}
c_{1,2}(t+1)=\\
\ups\int\int
f\left(\sqrt{\frac{[q_1(t)+\sig^2]
[q_2(t)+\sig^2]-c_{1,2}(t)^2}{q_2(t)+\sig^2}}
\xi_1+\frac{c_{1,2}(t)}{\sqrt{q_2(t)}}\xi_2+m_1(t)-\ta\right)f[\sqrt{q(t)+\sig^2}\xi_2+m_2(t)-\ta]
d\ga(\xi_1)d\ga(\xi_2)
\end{array}
\end{equation}

The proof is detailed in \cite{Moynot99}.

It is obvious now to infer the evolution of the mean quadratic
distance from the following square expansion.\\

\begin{prop}
The mean quadratic distance obeys the relation

\beq
d_{1,2}(t)=q_1(t)+q_2(t)-2c_{1,2}(t)+[m_1(t)-m_2(t)]^2
\eeq

\end{prop}

\subsubsection{Study of the special case of balanced inhibition} \label{SBalIn}

In order to show how the previous equations are used we shall
display the special case of balanced inhibition and excitation.
The study of the discrete time 1-dimensional dynamical system with
different parameters was addressed  in paper I.
See also (\cite{Cessac94} and \cite{Dauce98}) for more details.

We choose in the previous model the special case where
$\oups=0$. This choice simplifies considerably the evolution
study since $\forall t, m(t)=0$ and the recurrence over $q(t)$ is
autonomous. So we have just to study the attractors of a single
real function.

\begin{figure}
\begin{center}
\includegraphics[height=6cm,width=8cm,clip=false]{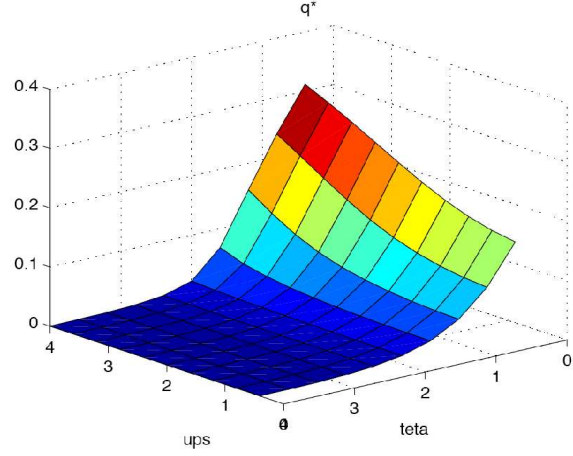}
\caption{Variations of the fixed point $q^*(\ups,\ta)$  as a function
of the network configuration parameters}
\label{fig:surfq}
\end{center}
\end{figure}

Moreover, the interpretation of $\oups=0$ is that there is a
general balance  in the network between inhibitory and excitatory
connections. Of course, the model is still far from biological
plausibility since the generic neuron is endowed both with
excitatory and inhibitory functions. In next section, the model with
several populations will be addressed. Nevertheless, the case
$\oups=0$ is of special interest.
In the limit of low noise, the mean-field dynamical system amounts to the recurrence
equation:
\begin{equation}\label{mfeq.eq}
q(t+1)=\ups\int{f[\sqrt{q(t)}\xi-\ta]^2d\ga(\xi)}
\end{equation}
we can scale $q(t)$ to $\ups$ and we obtain
\begin{equation}\label{mfeq2.eq}
q(t+1)=\int{f[\ups\sqrt{q(t)}\xi-\ta]^2d\ga(\xi)}=h_{\ups,\ta}[q(t)]
\end{equation}
\nid where the function $ h_{\ups,\ta}$ of $\RR^+$ into $\RR^+$ is
defined by
$$ h_{\ups,\ta}(q)=\int{f[\ups\sqrt{q(t)}\xi-\ta]^2d\ga(\xi)}$$
This function is positive, increasing  and tends to $0.5$
when $q$ tends to infinity. The recurrence (\ref{mfeq2.eq}) admits on
$\RR^+$ a single stable fixed point  $q^*(\ups,\ta)$.
This fixed point is increasing with $\ups$ and decreasing with
$\ta$. We represent in figure \ref{fig:surfq} the
diagram of the variations of function $q^*(\ups,\ta)$. It is
obtained from a numerical simulation with a computation of
$h_{\ups,\ta}$  by Monte-Carlo method.

Let us now consider the stability of the network dynamics by studying
the covariance and the mean quadratic distance evolution equation.
The covariance evolution equation (\ref{MFcovariance.eq}) in the
low noise limit and when $t\ra\infty$ amounts to
\begin{equation}\label{MFcovariance2.eq}\begin{array}{ll}
c(s+1,t+1)=\ups\int\int&
f\left(\sqrt{\frac{q^{*2}-c(s,t)^2}{q^*}}
\xi_1+\frac{c(s,t)}{\sqrt{q^*}}\xi_2-\ta\right)
f\left(\sqrt{q^*}\xi_2-\ta\right)
d\ga(\xi_1)d\ga(\xi_2)
\end{array}\end{equation}
Scaling the covariance with $\ups$ we obtain the recurrence
$$
c(s+1,t+1)=H_{\ups,\ta,q}[c(s,t)]
$$
with
\begin{equation}\label{MFcovariance3.eq}
H_{\ups,\ta,q}(c)=\int\int
f\left(\ups\sqrt{\frac{q^2-c^2}{q}}
\xi_1+\frac{c}{\sqrt{q}}\xi_2-\ta\right)
f\left(\ups\sqrt{q}\xi_2-\ta\right)
d\ga(\xi_1)d\ga(\xi_2)
\end{equation}

It is clear from comparing with equation (\ref{mfeq.eq}) that $q^*$
is a fixed point of $H_{\ups,\ta,q}$. To study the stability of
this fixed point, standard computation shows that

\begin{equation}\label{MFcovariance4.eq}
\frac{dH_{\ups,\ta,q^*}}{dc}(q^*)=\int
f'\left(\ups\sqrt{q^*}\xi_2-\ta\right)^2d\ga(\xi)
\end{equation}

Then, as it is stated in paper I, the condition
$\frac{dH_{\ups,\ta,q^*}}{dc}(q^*)\leq 1$ is a necessary and
sufficient condition for the stability of $q^*$. A detailed and
rigorous proof for $\ta=0$ is provided in \cite{Moynot99}. Then two
cases occur.

\begin{figure}
\begin{center}
\includegraphics[height=7cm,width=7cm,clip=false]{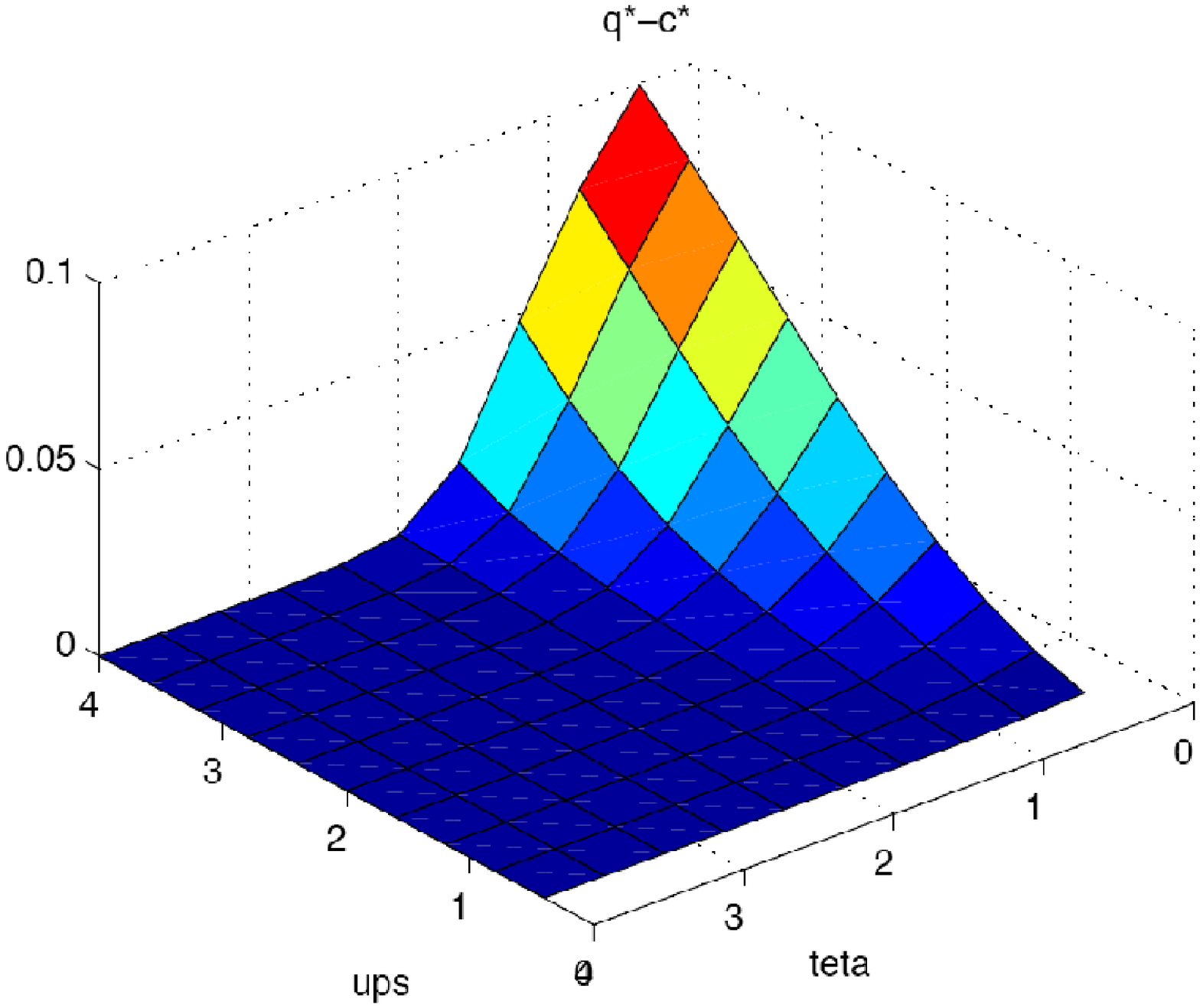}
\includegraphics[height=7cm,width=7cm,clip=false]{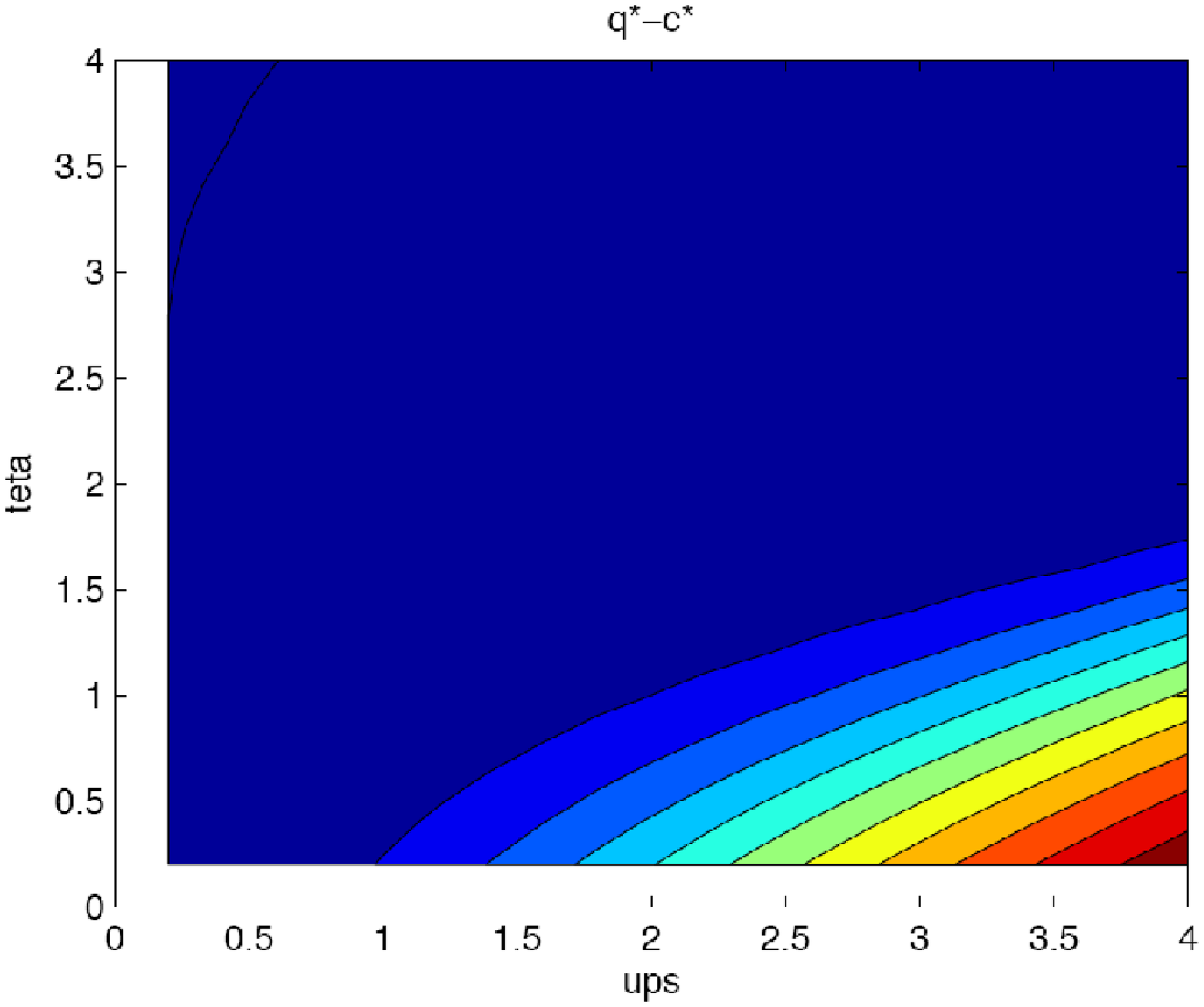}
\caption{Variations of $q^*-c^*$  as a function of the network
configuration parameters $\ups$ and $\ta$}
\label{fig:surfdist}
\end{center}
\end{figure}

\begin{itemize}
\item In the first case where $\frac{dH_{\ups,\ta,q^*}}{dc}(q^*)
\leq 1$,the stationary limit of $c(s+\tau,t+\tau)$ when
$\tau\ra\infty$ does not depend on $t-s$ and is
$c^*=q^*$. The stationary limit of the mean-field Gaussian process
is a random point. Its variance is increasing with $\ups$ and
decreasing with $\theta$.

\item In the second case where $\frac{dH_{\ups,\ta,q^*}}{dc}(q^*)>
1$ does not depend on $t-s$ when $t-s\neq 0$ and is equal to
$c^*<q^*$. The stationary limit of the Gaussian process is the sum
of a random point and of a white noise. From the dynamical system
point of view, this corresponds to a chaotic regime (with infinitely
many degrees of freedom). The signature of
chaos is given by the evolution of the mean quadratic distance. The
instantaneous covariance converges also towards $c^*$. Therefore
the mean quadratic distance converges towards a non zero limit,
which is independent of the initial condition distance. As shown in \cite{Cessac95}
 the transition from  fixed point to chaos is given by an explicit
equation which is the same as the equation of the De Almeida-Thouless line \cite{AT}
in spin-glasses models. The analogy between these two systems is further developed
in \cite{EPL,Cessac95}.

\end{itemize}
The figures \ref{fig:surfq} and \ref{fig:surfdist} shows the
evolution of $q^*$ and $q^*-c^*$ as a function of $\ups$ and $\ta$.
When $\ups$ is small, there is no bifurcation. When
$\ups$ is larger, a transition to chaos occurs when $\ta$ is
decreasing. When $\ups$ is growing, the transition to chaos
occurs for increasing $\ta$ values. Figure 31 of paper I
shows the interest of variation of input (which is
equivalent to threshold variation) allows to hold up the
occurence of chaos.

\section{MFT-based oscillation analysis in IF networks.} \label{MFOscIF}

In this section we would like to give an interesting application of
mean-field approaches for spiking neurons. It was developed in
\cite{BrunelHakim99}. This paper is part of a current of research which
studies the occurrence of synchronized oscillations
  in  recurrent spiking neural networks
\cite{Amit97CC,Amit97Net,Brunel00CNS}, in order to give
an account of spatio-temporal synchronization effects which are
observed in many situations in neural  systems
\cite{Gray94,SingerGray95,Buzsaki95,RitzSejnowski97}.

\subsection{IFRRNN continuous-time model}\label{IFRNNCont}

The model of \cite{BrunelHakim99} has continuous time. There is no
synaptic noise but the neurons are submitted to a random external
output. So, equation (\ref{IFdyn.eq}) has to be replaced by

\begin{equation}\label{IFdif2.eq}\left\{\begin{array}{ll}
u(t)<\ta&\Rightarrow \tau\du(t)=-u(t)+v_{net}(t)+v_{ext}(t)
\\
u(t-0)=\ta&\Rightarrow u(t+0)=\vta
\end{array}\right.\end{equation}
where \begin{itemize}
\item $\tau$ is the \textit{characteristic time} of the neuron,
\item $v_{net}$ is the synaptic input from the network,
\item $v_{ext}$ is the external input,
\item $\vta$ is the reset potential and $0<\vta<\ta$. Note
that $u(t-0)$ and $u(t+0)$ are respectively the left and right
limits of $u$ at firing time $t$. Thus, the refractory period is
assumed to be zero.
\end{itemize}
This model of continuous time neuron dynamics is introduced in
paper I, section 2.2.4.

Moreover, since the inputs are modeled by continuous-time stochastic
processes, equation (\ref{IFdif2.eq}) is a stochastic differential
equation of the type
\begin{equation}\label{IFdifstoch.eq}
\tau du(t)=-u(t)dt+dV_t
\end{equation}
with $dV(t)=dV_{ext}(t)+dV_{net}(t)$

Now we shall explicit these stochastic processes, in order to obtain
the Fokker-Planck equation of the network dynamics, in a mean-field
approximation.

\subsection{Modeling the external input}\label{ModExtI}
The network is a recurrent inhibitory network and we study its
reaction to random excitatory synaptic inputs. We suppose that in
the network each neuron receives excitations from $C_{ext}$ external
neurons connected via  constant excitatory synapses $J_{ext}$.
The corresponding external current is a Poisson process with emission
frequency $\nu_{ext}$.

Let us examine the effect of a superimposition of a large number $C$
of independent identically distributed low-rate $\nu$ Poisson processes.
Put
$$ \mc{I}(t)=J\sum_{i=1}^C \mc{N}_i(t)$$
where $\mc{N}_i(t)$ are i.i.d. Poisson processes with firing rate
$\nu$.
Then $I(t)$ is a stochastic process with independent stationary
increments such that $\EE(\mc{I}(t))=\mu t= JC\nu t$ and
$\Var(\mc{I}(t))=\sig^2 t=J^2C\nu t$.
Thus $\mu=JC\nu$ and $\sig=J\sqrt{C\nu}$.

We are interested in studying such processes when they reach the
firing threshold $\ta$ which is far larger than the elementary
increment $J$. In typical neural applications, $J=0.1$ mv and
$\ta=20$ mV. At this level, operating a classical time-space
rescaling, $\mc{I}(t)$ appears like a Gaussian process with
independent increments and same moments. We have
$$d\mc{I}(t)\sim\mu dt+\sig dB_t$$
where $(B_t)$ is the standard Brownian motion.
If we apply the subsequent to the external synaptic input we get the
following modeling in the limit of large size and low rate
$$dV_{ext}(t)=\mu_{ext} dt+\sig_{ext} dB(t)$$
with $\mu_{ext}=J_{ext}C_{ext}\nu_{ext}$ and
$\sig_{ext}=J_{ext}\sqrt{C_{ext}\nu_{ext}}$.

\subsection{Mean-field approximation of the internal input}\label{MFII}
In the framework of continuous-time modeling, the synaptic input
definition of $v_{net}$ for IF neuron $i$ which was, according to
equation (\ref{synapticdyn.eq}),
$$
v_i(\J,u)(t+1)=\sum_{j=1}^N J_{ij}x_j(t),
$$
has to be replaced by
\begin{equation}\label{SynCur}
v_i(\J,u)(t)=\tau \sum_{j=1}^N J_{ij} \sum_k \delta\left(t-T_j^k(u)
-D
\right)
\end{equation}
where
\begin{itemize}
\item $\del$ is the Dirac distribution,
\item $T_j^k(u)$ are the successive firing times of neuron $j$
during the network trajectory $u$,
\item $D$ is the synaptic transmission delay.
\end{itemize}

In the present study, the network is supposed to be sparsely connected. All
the connexion weights are equal to $-J$ as soon as they are non
zero. Each neuron is connected to $C$ neurons which are randomly
drawn among the network with  $C << N$ connections, where $C$ is a
fixed integer and $N$ is the total number of neurons. Another model
is considered further where the connection weights are independent
random variables equal to $-J$ with probability $\frac{C}{N}$ and to
0 else. We shall focus here on the first model.\\

In previous sections, mean-field approximation in the finite time set framework
consisted  in finding a fixed point for the
mean-field propagation operator $L$. Namely:
\begin{itemize}
\item approximating random vectors $v_i$ by Gaussian vectors with a probability distribution
$g_\mu$, where $\mu$ is a probability law on the individual neuron
potential trajectory space (finite-dimensional vector space)
\item finding $\mu$ as the probability law of the neuron
dynamical equation with this approximation for the synaptic
input.
\end{itemize}

\nid The mean-field approximation in \cite{BrunelHakim99} follows the
same logic. \\

The first step of the mean field approximation consists
for a given rate function $\nu$ in defining the non stationary
Gaussian process

\begin{equation}\label{Vnet.eq}
dV_{net}(t) = \mu_{net}(t)dt+\sigma_{net}(t) dB(t)
\end{equation}

where

\begin{itemize}
\item the drift $\mu_{net}$ is given by
\begin{equation}\label{munet.eq}
\mu_{net}(t)=-CJ\nu(t-D)\tau
\end{equation}

\item
and where the diffusion coefficient $\sig_{net}$ is given by
\begin{equation}\label{sigmat}
\sig_{net}(t)^2= J^2 C\nu(t-D)\tau
\end{equation}

\end{itemize}

\vspace{5mm}

The second step consists in considering the following diffusion with
"tunneling effect"
\begin{equation}\label{IFdiffus.eq}\left\{\begin{array}{ll}
u(t)<\ta&\Rightarrow \tau du(t)=-u(t)dt+dV_{net}(t)+dV_{ext}(t)
\\
u(t-0)=\ta&\Rightarrow u(t+0)=\vta
\end{array}\right.\end{equation}

The terminology ``tunneling effect'', referring to quantum mechanics,
is somewhat curious here. It has its roots in the following remark.
Whenever the membrane
potential reaches $\theta$ it is reset to $\vta$. If we interpret
eq. (\ref{IFdiffus.eq}) in the context of a random particle motion,
the ``particle'' is ``instantaneously transported'' from
the point $\ta$ to $\vta$. This analogy is not only formal.
The ``tunneling effect'' induces a specific behavior for the probability current
at the boundary $u=\theta$. In the present model, this current is directly related to the firing
rate (see next section).

\subsection{Fokker-Planck equation.} \label{FP}

\subsubsection{Closed form equation}
Note $p(u,t)$ the probability density of the solution $u(t)$ of
(\ref{IFdiffus.eq}).
Define
$$\begin{array}{l}
\mu(t)=\mu_{net}(t)+\mu_{ext}(t)
\\
\sig(t)=\sqrt{\sig_{net}(t)^2+\sig_{ext}(t)^2}
\end{array}$$
Then $p(u,t)$ is solution of the Fokker-Planck equation for diffusion
process for $u<\ta$ and $u\neq\vta$:
\begin{equation}\label{FokkerPlanck.eq}
\frac{\partial p}{\partial t}(u,t)=
\frac{\sig(t)^2}{2}\frac{\partial^2 p}{\partial u^2}(u,t) +
\frac{\partial}{\partial u}\left[(u-\mu(t))p(u,t)\right]
\end{equation}
The tunneling effect from $\ta$ to $\vta$ is taken into account in
the following boundary conditions
\begin{equation}\left\{\begin{array}{l}
p(\ta,t)=0
\\
\frac{\partial p}{\partial u}(\vta+0,t)=
\frac{\partial p}{\partial u}(\vta-0,t)+
\frac{\partial p}{\partial u}(\ta-0,t)
\end{array}\right.\label{BCFP.eq}\end{equation}

This corresponds to a re-injection of the outgoing
probability current $j(\theta,t)$ at $u=\vta$, 
where $j=\frac{\partial p}{\partial u}$. Thus $j(\vta+0,t)=j(\vta-0,t)+j(\theta,t)$.
The outgoing current (re-injected current) is, by definition,
the average firing rate defined by
\begin{equation}\label{FiringRate.eq}
\nu(t)=\frac{\partial p}{\partial u}(\ta-0,t)
\end{equation}

\subsubsection{Stationary solution}
It is easy to find the stationary solution of the previous
equation
$$\frac{\partial p}{\partial t}(u,t)=0$$
Suppose a given constant firing rate $\nu_0$, then set
\begin{equation}\label{Stat.eq1}\left\{\begin{array}{l}
\mu_0=-CJ\nu_0\tau + \mu_{ext}
\\
\sigma_0 =\sqrt{ C J^2 \nu_0 \tau + \sigma_{ext}^2}
\end{array}\right.\end{equation}

\nid and plug it into the differential second order equation
\begin{equation}\label{FokkerPlanck0.eq}
\frac{\sig_0^2}{2}\frac{d^2 p}{d u^2} +
\frac{d}{d u}\left[(u-\mu_0)p(u)\right]=0
\end{equation}
with the following boundary conditions
\begin{equation}\left\{\begin{array}{l}
p(\ta)=0
\\
\frac{d p}{d u}(\vta+0)=
\frac{d p}{d u}(\vta-0)+
\frac{d p}{d u}(\ta-0,t)
\end{array}\right.\label{BCFP0.eq}\end{equation}

One obtains easily the following stationary distribution
$$\left\{\begin{array}{ll}
\mbox{For}\ u<\vta,&
p(u)=\frac{2\nu_0}{\tau}e^{-y_u^2}
\int_{\ups_\vta}^{\ups_\ta}e^{y^2}dy
\\
\mbox{For}\ u\geq\vta,&
p(u)=\frac{2\nu_0}{\tau}e^{-y_u^2}
\int_u^{\upsilon_\ta}e^{y^2}dy
\end{array}\right.$$
where
$y_u=\frac{u-\mu_0}{\sig_0},y_\vta=\frac{\vta-\mu_0}{\sig_0}$
and
$y_\ta=\frac{\ta-\mu_0}{\sig_0}$

Then the normalization condition $\int_{-\infty}^\infty p(u)du=1$
allows to infer
\begin{equation}\label{Stat.eq2}
\frac{1}{\nu_0 \tau} = \int_{0}^{+\infty} e^{-y^2}
\left[\frac{e^{2 y_\theta y} - e^{2 y_\vta y}}{y} \right]dy
\end{equation}

The relations (\ref{Stat.eq1},\ref{Stat.eq2}) allows to compute
numerically $\nu_0$.  The equation (\ref{Stat.eq2}) can be
approximately solved in the situation  where the fluctuations
$\sigma_0$ are weak (i.e. $y_\ta>>1$ which means that the spiking
events are rare). In this case :

\begin{equation}\label{Stat.eq3}
\nu_0 \tau \approx \frac{y_\ta}{\sqrt{\pi}} e^{-y_\ta^2}
\end{equation}

This asymptotic expression can be compared to the escape probability
from the equation of motion of a particle in a parabolic potential
well $ {\cal V}$, with minimum  $\mu_0$, submitted to a Brownian
excitation
$$\tau dV_t = -(V-\mu_0)dt+\sigma_0 dB_t$$
The time rate to reach $V=\theta$ is thus given by the Arrhenius
time
$$\nu_0\tau\sim  e^{-y_\ta^2}$$

Numerical values of $\nu_0$ which are inferred from equations
(\ref{Stat.eq2}) and (\ref{Stat.eq3}) are compared in
\cite{BrunelHakim99} to the result of numerical simulations of
the network and there is a good agreement between theoretical
predictions and simulated firing rates.

\subsubsection{Stability analysis.}
The stability analysis for the stationary solution uses normal form
techniques similar to those described in paper I, but in an infinite
dimensional space. The Fokker-Planck equation is rescaled and expanded around
the steady-state solution. This intricate computation is fully
detailed in \cite{BrunelHakim99} . We simply focus to the
results.

\begin{figure}
\begin{center}
\includegraphics[height=7cm,width=7cm,clip=false]{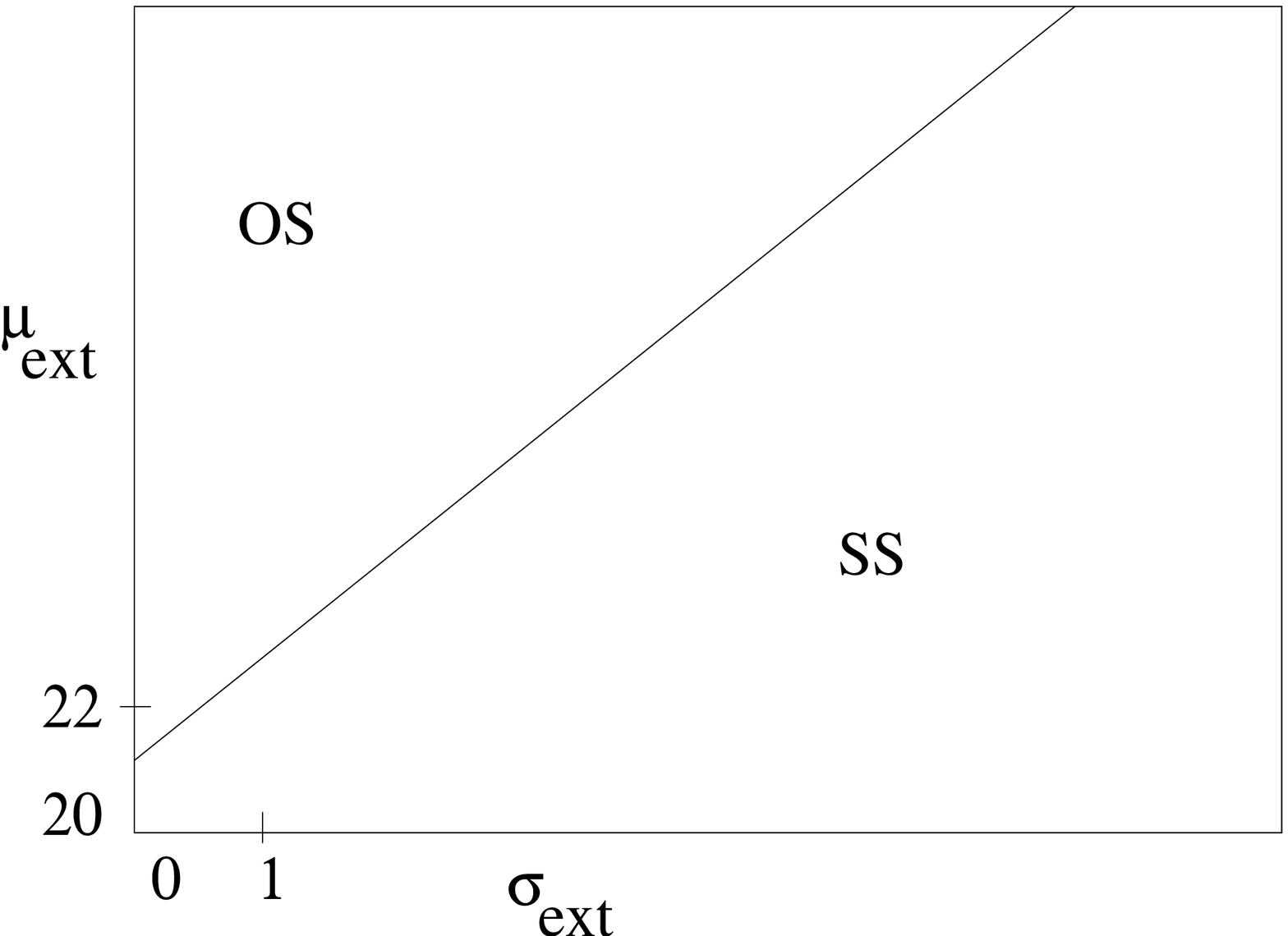}
\caption{Sketch of the bifurcation diagram of the model
(\ref{IFdif.eq},\ref{SynCur}) when varying the parameters $\mu_{ext},
\sigma_{ext}$  controlling the Poisson process of external
excitation. SS means Stationary State, while OS means Oscillatory
State. The solid line represents the instability line for $D=0.1
\tau$. (Drawn by hand from \cite{BrunelHakim99})
\label{fBifMapB1}}
\end{center}
\end{figure}

The Authors find that there is a bifurcation  of Hopf type for the
stationary solution. Thus, for a certain parameter  range, the system
exhibits synchronized oscillations of the neurons.  A sketch of the
bifurcation map is given in figure \ref{fBifMapB1} when varying the
parameters $\mu_{ext},\sigma_{ext}$  controlling the external
excitation.

One can see from that bifurcation diagram that the bifurcation
occurs when the drift of the external input is increasing. On the
opposite, an increase of the dispersion of the external input
stabilizes the steady state. If the external input consists in the
superposition of i.i.d. Poisson processes as it was detailed
above, then the increase of their common frequency $\nu_ext$
induces the occurrence of an oscillatory regime. There is still a
good agreement between the predictions of mean-field theory and the
results of simulations.

\subsection{Conclusion} \label{MFOscConc}

Thus, the conclusion is that in this model of a neural network with a
sparsely connected inhibitory integrate-and-fire neurons, submitted
to a external excitatory Poisson process, and emitting spikes
irregularly at a low rate, there is, in the thermodynamic limit, a
sharp transition between a regime where the average global is
constant, to a synchronized state where neurons are weakly
synchronized. The activity becomes  oscillatory when the inhibitory
feedback is strong enough. Note that the period of the global oscillations depends on the
synaptic transmission delay which cannot be neglected.

Finally, let us mention that the Authors performed a finite size analysis
of the model and found that global oscillations of \textit{finite coherence time}
generically exist \textit{above and below} the critical inhibition threshold.

\section{Appendix about probability theory}

This paper uses intensively some classical notations and
concepts coming from probability theory. The proofs are omitted but
sometimes the results follow from advanced results of this
theory. It is not possible to recall here the necessary
prerequisites. There are excellent books about probability theory
for physicists and engineers such as \cite{VanKampen92}. We just
want here to recall some notations and some results from
convergence theory. We have detailed the proof of the
"finite-time Girsanov theorem" since it is a crucial result for
the paper.

\subsection{Elementary Notations}
The classical and shortest point of view for considering random
phenomena from the 19th century is to consider a random variable
$x$ in a space $\E$ via its probability law on that space. 
All the moments  can be computed by integration
 over the probability law of the random
variable. For instance, if $\mu$ is the probability law of the
real random variable $x$, one has
\[
\EE(x)=\int_\E xd\mu(x)
\]
\[
\EE(x^2)=\int_\E x^2d\mu(x)
\]
and more generally for any bounded continuous function $\phi$ of $x$
\[
\EE[\phi(x)]=\int_\E \phi(x)d\mu(x)
\]
where $\EE$ is the\textit{ mathematical expectation} operator. The
expectation of any random variable
is a vector in a
topological vector space $\F$. The mathematical expectation
operator is linear.

Moreover, for a random vector $x \in \RR^d$ the
 expectation $E(x)\in\RR^d$ is defined by
\[
\forall i\in\{1,...,n\}, \{\EE(x)\}_i=\EE(x_i)=\int_{\RR^d}x_i
d\mu(x)
\]
and the symmetric $(d,d)-$covariance matrix is given by
\[
\mbox{Cov}(x)_{ij}=\EE(x_ix_j)-\EE(x_i)\EE(x_j)
\]
\nid where $\mu$ is the probability law of $x$.

Actually, this point of view cannot be used when we are obliged
to consider an infinite set of random variables or when we want to
operate a variable change. Hence, we are obliged to adopt a more
general point of view which was initiated by Kolmogorov in 1933.
This approach relies basically upon the consideration of a very
large state space $\Om$ which describes all the possible outcomes
or states of the world. Then a rich family $\A$ of subsets of
$\Om$ is defined such that all the random events of interest are
belonging to $\A$. Eventually a probability measure is defined on
$\A$ which associates to any random event $A\in\A$ its
probability $P(A)$. The triple $(\Om,\A,P)$ is called a
\textit{probability space}.

Later on, we shall have to work on infinite-dimensional space. So
let us fix a general framework

\begin{defi}
A Polish space $\F$ is a metric complete (every Cauchy sequence
converges) and separable (there is a countable dense subset)
space. The $\sigma-$algebra $\B$ of Borel subsets of A Polish
space $\F$ is the smallest $\sigma-$algebra that contains the
open sets.  Given a probability measure $\mu$ on the Borel
subsets of $\F$ it is possible to integrate any bounded
continuous function $\phi$ on $\F$ and the integral is noted
$\int_\F\phi(\xi)d\mu(\xi)$. The integral may be extended to a
wider class of functions. These functions are called
\underline{integrable}  with respect to $\mu$.
\end{defi}

In that new framework let us define random variables in $\F$.

\begin{defi}\label{law.def}
Let $(\Om,\A,P)$ be a probability space and $(\F,\B)$ a Polish
space endowed with its Borel $\sigma-$algebra. A \textit{random
variable} $x\in\F$ is a state function from $\Om$ into $\F$ such
that for any open set $B$ in $\F$, the subset of $\Om$ defined by
\[
(x\in B)=\{\om\in\Om\mbox{ such that }x(\om)\in\B\}
\]
belongs to $\A$ so its probability $P(x\in B)$ is well defined.

The \textit{probability law} of a random variable $x\in\F$ is the
probability law on $\F$ which associates to any Borel subset
$B\subset\F$ the probability $P(x\in B)$.
\end{defi}

The probability law of $x$ is noted $x.P$ or $P_x$. This
definition
stands for also for general measure than probability laws such as
volume measures. More generally, we have

\begin{defi}
Let $(\Om,\A,P)$ be a measure space and $x$ a mapping from $\Om$ to $\F$
such that
\[
\forall B\in\B, (x\in B)=\{\om\in\Om\mbox{ such that =
}x(\om)\in\B\}\in\A
\]
Then we define a measure on $(\F,\B)$ that is
noted $x.P$ or $P_x$ by
\[
B\in\B\rightarrow x.P(B)=P_x(B)=P(x\in B)
\]
This measure is called the \underline{image} of the measure $P$ by
the mapping $x$
\end{defi}

This definition is completed by the following transfer theorem
which shows that the mathe-matical expectation can be computed on
the state space $\Om$ or on the value space $\F$.

\begin{theo}
For any function $\phi$ defined on $\F$ and integrable for the
probability law $P_x$ we have
\[
\EE[\phi(x)]=\int_\Om \phi[x(\om)]dP(\om)=
\int_\F\phi(\xi)dP_x(\xi)
\]
\end{theo}

The transfer theorem is very useful in theory and in practice.
It allows to define the mathematical expectation of a random
variable without any ambiguity.

Kolmogorov's framework allows to define independent random variables by
the equivalent following properties

\begin{defi}
For $i\in\{1,...,n\}$ let $x_i\in\F_i$ be random variables, they
are said \underline{independent} if the law $P_x$ of the random
variable $x=(x_1,...,x_n)\in\F_1\times...\times\F_n$ is the
product of the $P_{x_i}$ which is expressed in the following
equivalent properties
\[ P(x\in B_1\times...\times B_n)=P_{x_1}(B_1)...P_{x_n}(B_n)   \]
\[ E[\phi_1(x_1)...\phi_n(x_n)]=E[\phi_1(x_1)]...E[\phi_n(x_n)]  \]
\end{defi}

\subsection{Density and Gaussian random vectors}
\begin{defi}
Let $(\Om,\A,m)$ a measure space and $h$ an integrable positive
function on $\Om$ such that $\int_\Om h(\om)dm(\om)=1$. Then we
can define a probability measure $Q$ on $(\Om,\A)$ by
\[
Q(A)=\int_\Om 1_A(\om)h(\om)dm(\om)
\]
$Q$ is said \underline{absolutely continuous} with respect to $m$,
$h$ is called the
\underline{density} of
$Q$ with respect to
$m$ and  we can compute the integral for $Q$ by using the formula
\[
\int_\Om \phi(\om)dQ(\om)=\int_\Om \phi(\om)h(\om)dm(\om)
\]
We write $\frac{dQ}{dm}(\om)=h(\om)$ or $dQ((\om)=h(\om)dm(\om)$
\end{defi}

Of course, the density functions are commonly used in elementary
probability. An important class of probability measures is the
Gaussian probability family.

\begin{defi}
Let $a\in\RR$ and $\sig^2\in\RR^+$. The Gaussian probability
measure $\ga=\N(a,\sig^2)$ is defined by its density with respect
to the Lebesgue measure $\la$ on $\RR$, which is
\[
\frac{d\ga}{d\la}(\xi)=\frac{1}{\sqrt{2\pi\sig^2}}
\exp\left[-\frac{(\xi-m)^2}{2\sig^2}\right]
\]
\end{defi}
Similarly, this definition can be extended to d-dimensional vector
  space and even to infinite-dimensional Hilbert space. Here, we
just need the following
\begin{defi}
Let $\ol{\ta}\in\RR^d$ and $K$ be a $d\times d$ symmetric positive
matrix, then there exists one and one only probability measure on
$\RR^d$, which is called the \underline{Gaussian probability}
$\ga=\N(\ol{\ta},K)$ such that if $\ga$  is the probability law
of the random vector $x\in\RR^n$ then
$\forall u\in\RR^d,$ the law of the random variable $u^t
x$\footnote{$u^t$ is the transpose of column vector  $u$, so $u^t x$ is
the scalar product of vectors $u$ and $x$} is
$\N(u^t\ol{\ta},u^tKu)$.
\end{defi}

\begin{prop}
Let $x$ be a random vector with regular Gaussian probability
$\ga=\N(\ol{\ta},K)$ then we have
\[\left\{\begin{array}{l}
\EE(x)=\int\xi d\ga(\xi)=\ol{\ta}
\\
\Cov(x)=\EE(xx^t)-\EE(x)\EE(x)^t=K
\end{array}\right.\]
\end{prop}

So a Gaussian law is completely determined by its expectation and
its covariance matrix.

\begin{defi}
With the previous notations, if $K$ is invertible, $\ga$ is said
to be \underline{regular} and the density  of $\ga$ with respect to the
Lebesgue measure
$\la$ is

\begin{equation}
\frac{d\ga}{d\la}(\xi)=\frac{1}{\sqrt{(2\pi)^n\mbox{Det}(K)}}
\exp\left[-\frac{(\xi-m)^tK^{-1}(\xi-m)}{2}\right]
\label{Gauss.eq1}
\end{equation}

\end{defi}
A common property of the Gaussian family is its stability by
linear transforms and translation. More precisely, we have

\begin{prop}\label{GaussLin.prop}
Let $x$ a Gaussian random vector which takes its value in the
vector space $E$ and $\La$ a linear mapping of $E$ into $F$.
Then $y=\La x$ is a Gaussian random vector in F and
\begin{equation}\left\{\begin{array}{l}
\EE(y)=\La \EE(x)
\\
\Cov(y)=\La \Cov(x)\La^t
\end{array}\right.\end{equation}
\end{prop}

\begin{prop}\label{GaussTrans.prop}
Let $x$ a Gaussian random vector which takes its value in the
vector space $E$ and $a\in E$. Then $y=x+a$ is a Gaussian
random vector in F and
\begin{equation}\left\{\begin{array}{l}
\EE(y)=\EE(x)+a
\\
\Cov(y)=\Cov(x)
\end{array}\right.\end{equation}
\end{prop}

\begin{coro}\label{Girsanovdis.prop1}
Let $x$ be a random vector with regular Gaussian probability
$\ga=\N(\ol{\ta},K)$ and let $a\in\RR^d$, then the law $\ga_a$ of
$x+a$ is the regular Gaussian law $\N(\ol{\ta}+a,K)$ and its
density with respect to $\ga$ can be written as follows
\begin{equation}
\frac{d\ga_a}{d\ga}(\xi)=
\exp\left[a^t K^{-1}(\xi-\ol{\ta})-\frac{1}{2}a^tK^{-1}a
\right]
\label{Girsanov.eq1}
\end{equation}
\end{coro}
\pr
The formula is checked using an easy and straightforward
computation from the expression of the Gaussian density
\cqfd\\

  It is interesting to note that it is possible to define Gaussian
probability on an infinite-dimensional vector space though it is
not possible to define Lebesgue measure. However, in that paper
we just use finite-dimensional Gaussian probabilities.  An
interesting property of the Gaussian measure, which is crucial in
this paper, is the following finite-dimensional version of the
Girsanov theorem\cite{Girsanov}.

\begin{theo}\label{Girsanovdis.theo}
Let $m_0$ a probability measure on $\RR^d$ and let $\N(\al,K)$ be
a Gaussian regular probability on $\RR^d$. Let $T$ a positive
integer and $\E_T=(\RR^d)^{\{0,...,T\}}$ the space of finite time
trajectories in $\RR^d$. Let $w$ a Gaussian random vector in
$\E_T$ with law $m_0\otimes\N(\al,K)^T$. Let
$\phi$ and $\psi$ two measurable applications of $\RR^d$ into
$\RR^d$. Then we define the random vectors $x$ and $y$ in $\E$ by

\[\left\{\begin{array}{l}
x_0=w_0\\x(t+1)=\phi[x(t)]+w(t+1)
\end{array}\right.\]
\[\left\{\begin{array}{l}
y_0=w_0\\y(t+1)=\psi[y(t)]+w(t+1)
\end{array}\right.\]

Let $P$ and $Q$ be the respective probability laws on $\E$ of $x$
and $y$, then $Q$ is absolutely continuous with respect to $P$ and
we have
\begin{equation}\label{GirsanovDiffA.eq}
\frac{dQ}{dP}(\eta)=\exp\sum_{t=0}^{T-1}\left\{\begin{array}{l}
-\frac{1}{2}\{\psi[(\eta(t)]-\phi[(\eta(t)]\}^t
K^{-1}\{\psi[(\eta(t)]-\phi[(\eta(t)]\}
\\
+\{\psi[(\eta(t)]-\phi[(\eta(t)]\}^tK^{-1}
\{\eta(t+1)-\al-\phi[\eta(t)]\}
\end{array}\right\}\end{equation}
\end{theo}
\pr
The proof is a recursion on $T$. It is easy to check
(\ref{GirsanovDiffA.eq}) for $T=1$.
To reduce the
expression let us write down
$$y_0^T=(y(0),...,y(T)),\eta_0^T=(\eta(0),...,\eta(T))$$
and
$$
\Ta_T(\eta_0^T)=\sum_{t=0}^{T-1}\left\{\begin{array}{l}
-\frac{1}{2}\{\psi[(\eta(t)]-\phi[(\eta(t)]\}^t
K^{-1}\{\psi[(\eta(t)]-\phi[(\eta(t)]\}
\\
+\{\psi[(\eta(t)]-\phi[(\eta(t)]\}^tK^{-1}
\{\eta(t+1)-\al-\phi[\eta(t)]\}
\end{array}\right\}$$
Suppose (\ref{GirsanovDiffA.eq}) is true up to $T$
and let us compute the density of $y$ up to $T+1$.
Let $h$ be a bounded continuous test function defined on
$\E_{t+1}$. We have by conditioning with respect to $y_0^T$
$$
\EE\left[h(y(T+1),y_0^T)\right]=
\int\EE\left\{h(w(T+1)+\psi[\eta(T)],\eta_0^T)\right\}
dQ(\eta_0^T)
$$
where the expectation is taken with respect to $w(T+1)$, which is
independent from $y_0^T$. Let us explicit the Gaussian law
$\N(\al,K)$ and use the recursion hypothesis:

$$
\begin{array}{ll}
\EE&\left[h(y(T+1),y_0^T)\right]=
\\&C_K
\int\int h(\om+\psi[\eta(T)],\eta_0^T)
\exp\left\{-\frac{1}{2}(\om-\al)^tK^{-1}(\om-\al)\right\} \exp\Ta_T(\eta_0^T)d\om dP(\eta_0^T)
\end{array}
$$
where $C_K$ is the classic normalization constant for the Gaussian
law. Then let us perform the translation
$\varpi=\om+\psi[\eta(T)]$, it gives
$$
\begin{array}{ll}
\EE&\left[h(y(T+1),y_0^T)\right]=
\\&C_K
\int\int h(\varpi,\eta_0^T)
\exp\left\{-\frac{1}{2}(\varpi-\al-\psi[\eta(T)])^tK^{-1}
(\varpi-\al-\psi[\eta(T)])\right\}

\exp\Ta_T(\eta_0^T)d\varpi dP(\eta_0^T)
\end{array}$$

To simplify notations let us write down
$\zeta_T=\psi[\eta(T)]-\phi[\eta(T)]$, we have
$$\begin{array}{ll}
\EE&\left[h(y(T+1),y_0^T)\right]=
\\&C_K
\int\int h(\varpi,\eta_0^T)
\exp\left\{-\frac{1}{2}(\varpi-\al-\phi[\eta(T)]+\zeta_T)^tK^{-1}
(\varpi-\al-\phi[\eta(T)]+\zeta_T)\right\}
\exp\Ta_T(\eta_0^T)d\varpi dP(\eta_0^T)
\end{array}$$
Let us develop the quadratic form in the exponential
$$\begin{array}{l}
-\frac{1}{2}(\varpi-\al-\phi[\eta(T)]+\zeta_T)^tK^{-1}
(\varpi-\al-\phi[\eta(T)]+\zeta_T)
\\=-\frac{1}{2}(\varpi-\al-\phi[\eta(T)])^tK^{-1}
(\varpi-\al-\phi[\eta(T)])
-\frac{1}{2}\zeta_T^tK^{-1}\zeta_T
+\zeta_T^tK^{-1}(\varpi-\al-\phi[\eta(T)])
\end{array}$$

So we have
$$\begin{array}{l}
\exp\left\{-\frac{1}{2}(\varpi-\al-\phi[\eta(T)]+\zeta_T)^tK^{-1}
(\varpi-\al-\phi[\eta(T)]+\zeta_T)\right\}
\\=\exp\left\{\frac{1}{2}(\varpi-\al-\phi[\eta(T)])^tK^{-1}
(\varpi-\al-\phi[\eta(T)])\right\}
\exp\left\{-\frac{1}{2}\zeta_T^tK^{-1}\zeta_T
+\zeta_T^tK^{-1}(\varpi-\al-\phi[\eta(T)])\right\}
\end{array}$$
We obtain a product of two exponentials. The first one combines
itself with $C_K d\varpi dP(\eta_O^T)$ to give $dP(\eta_O^{T+1})$;
the second one combines itself with $\exp\Ta_T(\eta_0^T)$ to give
$\exp\Ta_{T+1}(\eta_0^{T+1})$. So we get eventually
$$
\EE\left[h(y_0^{T+1})\right]=\int
h(\eta_0^{T+1})\exp\Ta_{T+1}(\eta_0^{T+1})dP(\eta_0^{T+1})
$$
\cqfd

\subsection{Convergence of random variables}
The definition of probability is based upon the law of large
numbers (LLN). This last result may be roughly formulated as
follows:\\

\textit{When $(x_n)$ is an independent \footnote{Such a sequence
is called an i.i.d. sequence}sequence of random variables with the same
probability law
$p$ with two first moments
$c=\int x dp(x)$ and
$k=\int x^2 dp(x)$ then the sequence of empirical averages
$\ol{x}_n=\frac{\sum_{k=1}^n x_k}{n}$
converges towards $c$.}\\

This statement is not precise. The convergence may have several
senses. Some useful convergence concepts in probability theory are
the convergence in law, the convergence in probability and the
almost sure convergence.
Let us recall their definition.

\begin{defi} Let $(x_n)$ and $x$ be random variables on a probability=20
space
$(\Om,\A,P)$. The sequence of random variables $(x_n)$ is said to
\begin{itemize}
\item
\underline{converge  in law} to $x$ if and only if for any continuous
bounded function $h$, $ \EE[h(x_N)]\ra \EE[h(x)] $
\footnote{An equivalent condition is the convergence of their
characteristic functions (or Fourier transforms):
$\forall t\in\RR, \EE(\exp(itx_n))\ra\EE(\exp(itx))$}
\item \underline{ converge in probability} to $x$ if and only if
\[
\forall \eps>0, P(\abs{x_n-x}\geq\eps)\ra 0
\]
\item\underline{converge  almost surely} to $x$ if and only if
\[
\exists N\subset\Om \mbox{ with }P(N)=0 \mbox{ such that }
\forall\om\notin N, x_n(\om)\ra x(\om)
\]
\end{itemize}
\end{defi}

These definitions are stronger and stronger. Almost sure
convergence implies convergence in probability which implies in
turn convergence in law. Most mean-field computations of
mean-field equations in random neural networks use the
convergence of Fourier transforms through a Laplace limit
integral ensuring convergence in law.

However, from the point of view of practitioners, almost
sure convergence is more pleasant because a single realization of
the sequence $(X_n)$ allows to check the convergence. To check
the weaker convergence statements, a lot of realizations of the
sequence are necessary.

Let us return to the law of large numbers. The convergence in
probability of the sequence $(\ol{x}_n)$ is specially easy to show
since
$\EE(\ol{x_n})=c$ and $\Var(\ol{x}_n)=\frac{k-c^2}{n}$.
Then one has just to write the Bienaym\'e-Tchebychev inequality
$$P(\abs{\ol{x}_n-c}\geq\eps)\leq\frac{k-c^2}{n\eps^2}$$
But this convergence is not strong enough to show the almost
sure convergence (the so-called \underline{strong} large number law).

\subsection{Large deviation principle}

\subsubsection{Cramer's theorem}

One way to obtain the strong law is to show that the convergence
in probability occurs much faster than it appears from
Bienaym\'e-Tchebychev inequality.

Actually the following theorem was obtained by Cramer in the late
30's:

\begin{theo}
Let $(x_n)$  sequence of i.i.d. random variables with probability
law $\mu$ such that $\int\xi d\mu(\xi)=\ol{\ta}$. Then we have
$$\forall a>\xi,\ \frac{1}{n}\log P(\ol{x}_n>a)\ra -I(a) $$ where
$$I(a)=\max_{p\in\RR} pa-E[\exp(px)]$$
\end{theo}

This theorem can be extended to more general settings. It is the
subject of \textit{large deviation} theory. Let us first consider
the case of finite-dimensional random vectors \cite{Kallenberg01}

The following proposition is easy to prove:
\begin{prop}
Let $\mu$ a probability law on $\RR^d$ such that for
all $p\in\RR^d$, $\La(p)=\log\EE[\exp(p^tx)]$ exists. The function
$\La$ is called the \textit{log-generating function} of $\mu$. We
define its \textit{Legendre transform} $\La^*$ on
$\RR^d$ as follows:
$$\La^*(a)=\sup_{p\in\RR^d}[(p^ta)-\La(p)]$$
then\begin{enumerate}
\item[a)] $\La^*$ is a convex function (with $\infty$ as a
possible value)
\item[b)] $\forall a\in\RR^d, \La^*(a)\geq 0$
\item[c)] $a=\int \xi d\mu(\xi)\Leftrightarrow\La^*(a)=0  $
\end{enumerate}
\end{prop}

\pr
a) is straightforward, since the supremum of convex functions is
convex

b) comes from Jensen's inequality.

c) comes from $\La(0)=1$

\cqfd

Then we can state the Cramer's theorem for i.i.d. sequence of
finite-dimensional random vectors:

\begin{theo}\textit{Cramer's theorem:}

Let $(x_n)$ be a sequence of i.i.d. random vectors with a
probability distribution $\mu$ according to the assumption and the
notations of the previous proposition. Then for any Borel subset
$B$ of $\RR^d$, we have
\begin{equation}\label{LDPdef.eq1}
-\inf_{a\in B^o}\La^*(a)\leq
\frac{1}{n}\ul{\lim}_{n\ra\infty}\log[\PP(\ol{x}_n\in B^o)]\leq
\frac{1}{n}\ol{\lim}_{n\ra\infty}\log[\PP(\ol{x}_n\in\ol{B})]\leq
-\inf_{a\in\ol{B}}\La^*(a)
\end{equation}
where $B^o$ is the interior set of $B$ (the greatest open subset
of $B$) and $\ol{B}$ is the closure of $B$ (the smallest closed
extension of $B$).
\end{theo}

A consequence of Cramer's theorem is that for any closed subset
$F$ in $\RR^d$ such that $\inf_{a\in \ol{B}}\La^*(a)>0$,
$\PP(\ol{X}_n\in F)$ goes to $0$  exponentially fast when
$n\ra\infty$ and that the rate of convergence depends only on the
value of $\La^*$ at the point of $F$ where $\La^*$ reaches its
minimum. This point is called the dominating point. For regular
probability distributions where $\La^*$ is strictly convex,
defined and continuous around $\ol{\ta}=\EE(x)$, the exponential
decay of finite deviations from the expectation (large
deviations) and the strong law of large numbers are easy
consequences.

\subsubsection{Large deviation principle in an abstract
setting}

The convergence with an exponential rate is a general
situation,  which is characterized in the following general
definitions:

\begin{defi} Let $\E$ be a Polish space and $I$ be a lower
semi-continuous function of $\E$ into $[0,\infty]$. $I$ is called
a \textit{rate function}. If $I$ possesses the property of compact
level set, i.e.
\[
\forall\eps>0,\{x\in\E\mbox{ such that }I(x)\leq\eps\}
\mbox{ is compact}
\]
then $I$ is called a \textit{good rate function}.
\end{defi}

\begin{defi}\label{LDPA}
Given a rate function $I$ on a Polish space $\F$ and a sequence of
probability measures $Q_n$ on $\F$, if for any Borel subset $B$
of $\F$, \begin{itemize}
\item
$(Q_n)$ satisfies the \textit{large deviation minoration
on open sets} if
\begin{equation}\label{LDPdef.eq2}
\forall O, \mbox{ open set in }\F,-\inf_{\xi\in O}I(\xi)
\leq\frac{1}{n}\ul{\lim}_{n\ra\infty}\log[Q_n(O)]
\end{equation}
\item
$(Q_n)$ satisfies the \underline{large deviation majoration
on compact sets} if
\begin{equation}\label{LDPdef.eq3}
\forall K, \mbox{ compact set in }\F,
\frac{1}{n}\ol{\lim}_{n\ra\infty}\log[Q_n(K))]
\leq-\inf_{\xi\in K}I(x)
\end{equation}
\item
$(Q_n)$ satisfies the \underline{large deviation majoration
on closed sets} if
\begin{equation}\label{LDPdef.eq4}
\forall C, \mbox{ closed set in }\F,
\frac{1}{n}\ol{\lim}_{n\ra\infty}\log[Q_n(C))]
\leq-\inf_{\xi\in C}I(x)
\end{equation}
\item
If $(Q_n)$ checks the large deviation minoration for open sets
and the large deviation majoration for compact sets we say that
$(Q_n)$ satisfies the \underline{large deviation principle (LDP) }
with rate function $I$.
\item
If $(Q_n)$ checks the large deviation minoration for open sets
and the large deviation majoration for closed sets we say that
$(Q_n)$ satisfies the \underline{full large deviation principle}
with rate function $I$.
\item $(Q_n)$ is said \underline{tight} if for all $\eps>0$, it exists
a compact subset $K$ of $\F$ such that $Q_n(^cK)<\eps$. If
$(Q_n)$ is tight and checks a LDP, it satisfies the full LDP for
the same rate function.
\end{itemize}
\end{defi}
The same definitions stand for a sequence of random elements
in $\F$ if the sequence of their probability laws checks the
respective majorations.

A simpler way to state that $(Q_n)$ satisfy the full large
deviation principle with rate function
$I$ is to write that
\begin{equation}\label{LDPdef.eq5}
-\inf_{\xi\in B^o}I(\xi)\leq\frac{1}{n}
\ul{\lim}_{n\ra\infty}\log[Q_n(B)]
\leq\frac{1}{n}\ol{\lim}_{n\ra\infty}\log[Q_n(B))]
\leq-\inf_{\xi\in\ol{B}}I(x)
\end{equation}

Actually, the scope of Cramer's theorem may be widely extended
and a full large deviation principle is checked for the empirical
mean of any i.i.d. random sequence in a Polish space under mild
assumptions on the existence of the log-generating function
\cite{Dembo93}. The rate function of this LDP is the
Legendre transform of the log-generating function.

\subsubsection{Varadhan theorem and Laplace principle}
An equivalent functional formulation of the full large deviation
principle is due to Varadhan and is called by Dupuis and Ellis
the  \textit{Laplace principle} (\cite{DupuisEllis97}).

\begin{defi}
Let $I$ be a good rate function on the Polish space $\F$. The
random sequence $(x_n)$ in $\F$ is said to satisfy the
\underline{Laplace principle} with  rate function $I$ if for any
continuous bounded function $h$ on $\E$ we have
\[
\lim_{n\ra\infty}\frac{1}{n}\log \EE\{\exp[-nh(x_n)]\}
=-\inf_{\xi\in\F}\{h(\xi)+I(\xi)\}
\]
\end{defi}

This approach is called by the authors of  (\cite{DupuisEllis97})
the \textit{weak convergence approach} to the theory of large
deviations. The equivalence of the two approaches (Laplace
principle for good rate functions and full large deviation
principle with good rate functions) are expressed in a theorem of
Varadhan and its converse. Their proofs are in
(\cite{DupuisEllis97}). Handling continuous bounded test
functions may be more practical than dealing with open and closed
sets. In particular, it is very easy to show the following
transfer theorem for the LDP principle when the law is changed.

\begin{theo}\label{Varadhan.theo}
Let $P_n$ and $Q_n$ two sequences of probability measures on the
Polish space $\F$, let I be a good rate function on $F$ and let
$\Ga$ a continuous function on $F$ such that
\begin{enumerate}
\item[(a)] $Q_n<<P_n$ and $\frac{dQ_n}{dP_n}(\xi)=\exp n\Ga(\xi)$,
\item[(b)] $(P_n)$ satisfies a full large deviation principle
with rate function $I$,
\item[(c)] $I-\Ga$ is a good rate function,
\end{enumerate}
then $(Q_n)$ satisfies a full large deviation principle
with rate function $I-\Ga$.
\end{theo}
\prtheo
Using the weak large deviation approach and the strong hypothesis
of the theorem, the proof is quite formal. Let $h$ be any
continuous bounded test function on $\F$, from hypothesis (c)
and
\cqfd

\subsection{Convergence of random measures}
Let us have a second look at the law of large numbers. Since this
law claims the convergence on the sequence of empirical averages
$\frac{1}{n}\sum_{k=1}^n f(x_k)$ over any bounded continuous test
function $f$ we are lead to consider the empirical measure of a
sample.

\begin{defi}\label{empiric.def}
Let $\xi=(\xi_1,...,\xi_n)\in\RR^{nd}$ a sequence of $n$ vectors of=20
$\RR^d$.
We associate to $\xi$ the following probability measure
$\mu_\xi\in\Pp(\RR^d)$
\[
\mu_\xi=\frac{1}{n}\sum_{k=1}^n \del_{\xi_k}
\]
$\mu_xi$ is called the\underline{ empirical measure} associated to $\xi$.
\end{defi}

This definition says that if $A$ is a Borel subset of $\F$ then
$\mu_N(x)(A)$ is the fraction of neurons which state trajectory
belong to $A$. More practically, if $\phi$ is any test
continuous function on $\E$, it says that
\[
\int_\E\phi(\eta)d\mu_N(u)(\eta)=
\frac{1}{N}\sum_{i=1}^N\phi(u_i)
\]

With this definition, the convergence for each continuous bounded
test function $f$ of $\frac{1}{n}\sum_{k=1}^n f(x_k)$ towards
$\int f(\xi)d\mu(\xi)$ is exactly the narrow convergence of the
sequence $\mu_{x_n}$ towards $\mu$.

The set $\Pp(\RR^d)$ of probability measure on $\RR^d$ is a
convex subset of the functional vector space $\M^1(\RR^d)$ of
bounded measures on $\RR^d$. We endow  $\Pp(\RR^d)$ with the
\textit{narrow topology} for which $\mu_n\ra\mu$ if and only if
for all continuous and bounded test function
$f\in\C^b(\RR^d),\int fd\mu_n\ra\int fd\mu$.
$\Pp(\RR^d)$ is a Polish state for this topology.

So instead of considering the random variable $x$ which takes its
values in $\RR^d$, we consider the random variable $\del_\xi$
which takes its values in the Polish space $\Pp(\RR^d)$. If
$(x_k)$ is an i.i.d. sequence in
$\RR^d$ with probability law $\mu$, then $\del_{x_i}$ is an i.i.d.=20
sequence
in $\Pp(\RR^d)$ and its empirical mean is just $\mu_{(x_1,...,x_n)}$ the
empirical measure of an i.i.d. sample of size $n$. That means that=20
Cramer's
theorem  extension to Polish spaces may be applied.
This theorem is known as \textit{Sanov theorem}.

Let us first recall the definition of the relative entropy with
respect to a probability measure $\mu$ on $\RR^d$.

\begin{defi}\label{crossentropy.defi}
Let $\mu$ be a probability measure on $\RR^d$. We define a convex
function  $\nu\in\Pp(\RR^d)\ra  I(\nu,\mu)\in\RR$ by:
\begin{equation}
\left\{\begin{array}{l}
I(\nu,\mu)=\int \log \frac{d\nu}{d\mu}(\xi) d\nu(\xi)
\\
I(\nu,\mu)=\infty \mbox{ else}
\end{array}\right.
\end{equation}
This function is called the \underline{relative entropy} with
respect to $\mu$
\end{defi}

then we may state the Sanov theorem \cite{Ellis85},
\cite{Dembo93}
\begin{theo}\label{Sanov.theo}
The sequence of empirical measure $\mu_n$ which are associated to
size $n$ i.i.d. sample of a probability law on $\RR^d$ satisfy a
full LDP with the \underline{relative entropy} with respect to $\mu$
as the rate function.
\end{theo}

\pagebreak

\bibliographystyle{plain}
\bibliography{biblioDYNN}

\ed